# Modeling tau transport in the axon initial segment

Ivan A. Kuznetsov(a), (b) and Andrey V. Kuznetsov(c)

(a)Perelman School of Medicine, University of Pennsylvania, Philadelphia, PA 19104, USA

(b)Department of Bioengineering, University of Pennsylvania, Philadelphia, PA 19104, USA

(c)Dept. of Mechanical and Aerospace Engineering, North Carolina State University,

Raleigh, NC 27695-7910, USA; e-mail: avkuznet@ncsu.edu

## Abstract

By assuming that tau protein can be in seven kinetic states, we developed a model of tau protein transport in the axon and in the axon initial segment (AIS). Two separate sets of kinetic constants were determined, one in the axon and the other in the AIS. This was done by fitting the model predictions in the axon with experimental results and by fitting the model predictions in the AIS with the assumed linear increase of the total tau concentration in the AIS. The calibrated model was used to make predictions about tau transport in the axon and in the AIS. To the best of our knowledge, this is the first paper that presents a mathematical model of tau transport in the AIS. Our modeling results suggest that binding of free tau to MTs creates a negative gradient of free tau in the AIS. This leads to diffusion-driven tau transport from the soma into the AIS. The model further suggests that slow axonal transport and diffusion-driven transport of tau work together in the AIS, moving tau anterogradely. Our numerical results predict an interplay between these two mechanisms: as the distance from the soma increases, the diffusion-driven transport decreases, while motor-driven transport becomes larger. Thus, the machinery in the AIS works as a pump, moving tau into the axon.

**Highlights**

- Transport of tau protein in an axon initial segment (AIS) was investigated.
- In most of the axon, tau motor-driven fluxes are independent of distance from the soma.
- A negative gradient of free tau causes diffusion-driven flux of free tau into the AIS.
- Both motor-driven and diffusion-driven transport move tau anterogradely within the AIS.
- Deeper into the AIS, tau switches from diffusion-driven to motor-driven transport.

## 1. Introduction

The axon initial segment (AIS) is a region in the beginning of the proximal axon that, due to a high density of ion channels, serves as an action potential initiation site [1]. The special structure of the AIS is due to a high concentration of Ankyrin G (AnkG), which is a master scaffolding protein that is responsible for the localization, immobilizing, and assembly of a highly specialized protein network responsible for AIS functionality [2-5].

The AIS is between 20 to 60 μm in length. Usually, the start of the segment is located immediately following the axon hillock [1,6]. AIS morphology differs from the rest of the axon as it is characterized by the presence of microtubule (MT) bundles (fascicles), a thin undercoat lining of the plasma membrane, and the lack of ribosomes [7,8].

Investigating transport of various cargos in the AIS is important because the disruption of the AIS may lead to neurotoxicity. One of axonal proteins that may be affected by AIS disruption is tau. Experiments with tau-knockout mice suggest that tau is required for normal neuronal function [9,10]. However, tau is not commonly known for its normal biological functions, which are still not fully understood, but rather for its ability to aggregate in a number of neurodegenerative diseases, called tauopathies [11-13]. The most common tauopathy is Alzheimer's disease (AD) [14]. Neurofibrillary tangles (NFTs), one of the hallmarks of AD, consist of deposits of hyperphosphorylated misfolded tau protein in the brain. NFTs are found many years before the onset of clinical symptoms in AD [15,16]. It should be noted that unlike other tauopathies, AD is also characterized by aggregation of a second protein, amyloid β [17-19]. It is believed that progressive accumulation of toxic forms of tau and amyloid β may lead to synaptic loss in AD [20].

Due to failures of phase three clinical trials of various anti-amyloid therapies [21], interest in investigating the role of tau in AD has increased [22]. An understanding of kinetics and dynamics of tau transport would be beneficial for understanding the pathophysiology of AD and other tauopathies. The behavior of tau in the AIS is of particular interest, because malfunctioning of tau axonal transport is one of early signs of AD.

Ref. [23] reported that in the AIS, a large fraction of tau monomers, rather than being associated with individual MTs, are associated with MT bundles. They noted that the kinetics of tau interaction with MT bundles and individuals MTs is different. This explains why values of kinetic constants used in our model to describe tau interaction with MTs in the AIS are different from those in the rest of the axon.

Ref. [24] discovered a tau diffusion barrier (TDB) in the AIS. Experimental results reported in [24] show that the location of TDB coincides with the location of the axon initial segment (AIS). For our model development, it is important to note that, although the TDB colocalizes with the AIS, this barrier depends on MTs and is independent of F-actin, unlike the AIS [24]. Ref. [24] compared the effect of TDB on tau transport with a diode, since a TDB supposedly has low resistance to anterograde flux of tau and high resistance to retrograde flux of tau. The mechanism of TDB appears to be dependent on MTs rather than on AnkG, a marker for the AIS. Due to the presence of a TDB, tau cannot move out of the axon.

In this paper, we developed a model of tau transport in the axon and in the AIS. The model was fit using published experimental data in the axon. For simplicity, because we found the model was not very sensitive to the tau distribution in the AIS, we fit the model in the AIS assuming a linear distribution of tau concentration in the AIS. We then used the calibrated model to gain some insight about tau transport in the AIS.

The AIS can be viewed as a connector between the soma and the axon (Fig. 1). In doing this work, we asked whether slow axonal transport of tau, which moves tau due to its interaction with molecular motors that run along MTs [25,26], is efficient enough to move tau from the AIS to the distal axon. We also aim at clarifying the role of diffusion in tau transport in the AIS. In particular, ref. [27] suggested that slow axonal transport counteracts diffusion in the AIS, because the tau concentration in the axon is greater than in the soma. We used the model to clarify the role and direction of diffusion-driven transport of tau in the AIS.

Our goal is to understand how tau protein can be depleted from the somatodendritic compartment and enriched in the axonal compartment [28]. It is important to understand why in healthy neurons diffusion does not result in a net retrograde tau transport and does not lead to tau redistribution so that its concentration would be the same in both the axon and soma.

We also asked whether the retrograde barrier for tau in the AIS is really a diode, as suggested in [24], and allows only for a unidirectional (anterograde) flux of tau. Alternatively, it is possible that the AIS is permeable to tau in both retrograde and anterograde directions, and the anterograde flux of tau simply exceeds the retrograde flux, as is suggested in [27].

## 2. Materials and models

### 2.1. Model of tau protein transport in the axon and the AIS

We treated the axon as a composite domain [29,30] consisting of two regions, the AIS (region 1) and the rest of the axon (region 2); hereafter we drop the words “rest of” and simply call region 2 “the axon” (Fig. 1). To simulate tau transport in the axon and the AIS, we used a model developed in [31,32]. The model in [31] was developed to simulate tau transport in the axon only, and in order to extend this model to the AIS we used the following experimental observations.

Tau does not accumulate in the AIS, which suggests that tau is not trapped by MT fascicles in the AIS [24]. This justifies neglecting the transient terms in our model. Axonal vesicles can freely pass the AIS without slowing down. This finding does not support the existence of an F-active-based filter in the AIS cytoplasm, but rather supports regulated active transport of various axonal cargos through the AIS [33]. Therefore, we assumed that in the AIS, tau is driven by the same physical mechanisms as in the axon, and hence, tau transport in the AIS is described by the same equations as in the axon. However, due to the unique MT organization in the AIS (fasciculated MTs) [34], it is reasonable to expect that there are two different sets of kinetic constants ($\gamma$s) in the AIS and in the rest of the axon. These kinetic constants describe tau interactions with MTs and tau transitions between different kinetic states. It is common to use the same model with different sets of kinetic constants in different regions of the axon. Ref. [35], for example, used different sets of constants to describe neurofilament transport in the nodes of Ranvier and in the internodes.

Our model is a cargo-level model, rather than a motor-level model [36,37]. Our goal is to simulate the concentration of the cargo (which is tau in our case), rather than cargo interactions with motors, which is not being modeled here. Our model simulates tau in seven kinetic states, which are summarized in Fig. 2. Since an axon is particularly long in one direction, the tau concentrations in the axon and AIS are represented by the linear number densities of tau monomers in various kinetic states ($\mu m^{-1}$).

The main mechanisms of tau transport in the axon and AIS are active, motor-driven transport (slow axonal transport) [25,26,38], diffusion in the cytosol [23,39,40], and one-dimensional diffusion along the MT lattice [41]. Slow axonal transport is characterized by short quick runs and long pauses [36,42].

Active anterograde transport of tau is due to its binding to kinesin [26], while active retrograde transport of tau may be explained by tau’s binding to dynein [43]. Retrogradely transported tau monomers may also be moved by kinesin motors that are attached to immobile MTs. This happens because if the motors are fixed, their action leads to retrograde motion of small MT fragments that may have an attached tau monomer [28]. For the completeness of the paper, we briefly restate the governing equations developed in [31].

A Cartesian coordinate directed along the axon is denoted $x^*$. We defined two regions: region 1 ( $-L_{AIS}^* \le x^* \le 0$), which corresponds to the AIS, and region 2 ( $0 \le x^* \le L^*$), which corresponds to the axon (Fig. 1). We will solve model equations in both regions 1 and 2 and then match the solutions through the boundary conditions at the interface, $x^* = 0$.

The conservation of tau in motor-driven states leads to:

$$-v_a^* \frac{\partial n_{a(i)}^*}{\partial x^*} - \gamma_{10(i)}^* n_{a(i)}^* + \gamma_{01(i)}^* n_{a0(i)}^* = 0, \qquad (1)$$

$$v_r^* \frac{\partial n_{r(i)}^*}{\partial x^*} - \gamma_{10(i)}^* n_{r(i)}^* + \gamma_{01(i)}^* n_{r0(i)}^* = 0, \qquad (2)$$

where ($i$=1) corresponds to the AIS and ($i$=2) corresponds to the axon; $n_a^*$ is the concentration of on-track tau monomers moving along MTs anterogradely, propelled by molecular motors; $n_r^*$ is the concentration of on-track tau monomers moving along MTs retrogradely, propelled by molecular motors; $n_{a0}^*$ is the concentration of pausing on-track tau monomers that are still associated with molecular motors and can resume their anterograde motion; $n_{r0}^*$ is the concentration of pausing on-track tau monomers that are still associated with molecular motors and can resume their retrograde motion. Model parameters are defined in Tables S1 and S2 in the Supplemental Materials.

The first terms on the left-hand sides of Eqs. (1) and (2) simulate active (motor-driven) tau transport. The other terms in Eqs. (1) and (2) simulate tau transitions between kinetic states in which tau is actively transported and kinetic states in which tau is pausing (Fig. 2).

Tau can enter or leave the pausing states only via transitions to/from other kinetic states (Fig. 2). Stating the conservation of tau in the pausing states gives the following equations:

$$-\left(\gamma_{01(i)}^* + \gamma_{ar(i)}^* + \gamma_{off,a(i)}^*\right) n_{a0(i)}^* + \gamma_{10(i)}^* n_{a(i)}^* + \gamma_{ra(i)}^* n_{r0(i)}^* + \gamma_{on,a(i)}^* n_{free(i)}^* = 0, \qquad (3)$$

$$-\left(\gamma_{01(i)}^* + \gamma_{ra(i)}^* + \gamma_{off,r(i)}^*\right) n_{r0(i)}^* + \gamma_{10(i)}^* n_{r(i)}^* + \gamma_{ar(i)}^* n_{a0(i)}^* + \gamma_{on,r(i)}^* n_{free(i)}^* = 0, \qquad (4)$$

where $n_{free}^*$ is the concentration of free (off-track) tau monomers in the cytosol.

Stating the conservation of free (cytosolic) tau gives the following equation:

$$D_{free}^* \frac{\partial^2 n_{free(i)}^*}{\partial x^{*2}} + \gamma_{off,a(i)}^* n_{a0(i)}^* + \gamma_{off,r(i)}^* n_{r0(i)}^* - \left(\gamma_{on,a(i)}^* + \gamma_{on,r(i)}^* + \gamma_{free\to st(i)}^* + \gamma_{free\to dif(i)}^*\right) n_{free(i)}^* + \gamma_{st\to free(i)}^* n_{st(i)}^*$$

$$+\gamma^*_{dif\to free(i)} n^*_{dif(i)} - \frac{n^*_{free(i)} \ln(2)}{T^*_{1/2}} = 0\,, \tag{5}$$

where $n^*_{st}$ is the concentration of stationary tau monomers bound to MTs, no association with motors; and $n^*_{dif}$ is the concentration of tau monomers diffusing along MTs, no association with motors.

In the free state tau can be transported by diffusion (simulated by the first term in Eq. (5)). The last term in Eq. (5) describes tau degradation. As a soluble protein, tau is degraded by an ubiquitin-proteasome system. Tau is also degraded through autophagy [9,44,45]. The decay of free tau is assumed to be proportional to the tau concentration in the free (cytoplasmic) state. The rest of the terms in Eq. (5) simulate tau transitions to/from other kinetic states (Fig. 2). Results obtained in [27] show that tau can freely diffuse between the axon and soma, which justifies the use of Eq. (5) not only for the axon, but also for the AIS.

While free tau can diffuse in any direction, a portion of tau attached to MTs can diffuse along the MTs, although with a much smaller diffusivity than free tau. Stating the conservation of a sub-population of MT-bound tau that can diffuse along the MTs leads to the following equation:

$$D^*_{mt} \frac{\partial^2 n^*_{dif(i)}}{\partial x^{*2}} - \left(\gamma^*_{dif\to free(i)} + \gamma^*_{dif\to st(i)}\right) n^*_{dif(i)} + \gamma^*_{free\to dif(i)} n^*_{free(i)} + \gamma^*_{st\to dif(i)} n^*_{st(i)} = 0\,. \tag{6}$$

While the first term in Eq. (6) simulates tau diffusion, the rest of the terms simulate tau transition to/from other kinetic states (Fig. 2).

Part of tau attached to MTs is stationary [41], and stating tau conservation in the stationary kinetic state leads to the following equation:

$$-\left(\gamma^*_{st\to free(i)} + \gamma^*_{st\to dif(i)}\right) n^*_{st(i)} + \gamma^*_{free\to st(i)} n^*_{free(i)} + \gamma^*_{dif\to st(i)} n^*_{dif(i)} = 0\,. \tag{7}$$

All terms in Eq. (7) simulate tau transitions between different kinetic states (Fig. 2).

We found the total concentration of tau protein by calculating the sum of tau concentrations in all seven kinetic states (Fig. 2):

$$n^*_{tot(i)} = n^*_{a(i)} + n^*_{r(i)} + n^*_{a0(i)} + n^*_{r0(i)} + n^*_{free(i)} + n^*_{st(i)} + n^*_{dif(i)}\,. \tag{8}$$

In six out of seven kinetic states tau is attached to MTs. We then calculated the percentage of MT-bound tau by taking the ratio of the sum of tau concentrations in all kinetic states, except for $n^*_{free}$, to the total concentration of tau:

$$\%\text{bound} = \frac{n^*_{a(i)} + n^*_{r(i)} + n^*_{a0(i)} + n^*_{r0(i)} + n^*_{st(i)} + n^*_{dif(i)}}{n^*_{tot(i)}}\left(100\%\right). \tag{9}$$

The total flux of tau is the sum of tau fluxes due to the action of anterograde and retrograde motors and due to diffusion of free and MT-bound tau, respectively:

$$j^*_{tot(i)} = v^*_a n^*_{a(i)} - v^*_r n^*_{r(i)} - D^*_{free}\frac{\partial n^*_{free(i)}}{\partial x^*} - D^*_{mt}\frac{\partial n^*_{dif(i)}}{\partial x^*}. \tag{10}$$

Following [46], we found the average tau velocity as the ratio of the total tau flux to the total tau concentration:

$$v^*_{av(i)}\left(x^*\right) = \frac{j^*_{tot(i)}}{n^*_{tot(i)}}. \tag{11}$$

For region 2 ( $0 \le x^* \le L^*$ ) Eqs. (1)-(7) were solved subject to the following boundary conditions:

$$\text{At } x^* = 0: \quad j^*_{tot}\Big|_{x=0^+} = j^*_{tot,x=0}, \quad n^*_{free}\Big|_{x=0^+} = n^*_{free,x=0}, \quad n^*_{tot}\Big|_{x=0^+} = n^*_{tot,x=0}. \tag{12a,b,c}$$

At the axon terminal, we imposed the following boundary conditions:

$$\text{At } x^* = L^*: \quad \left.\frac{dn^*_{free}}{dx^*}\right|_{x=L} = 0, \quad n^*_{tot}\Big|_{x=L} = n^*_{tot,55}, \quad j^*_{tot} = j^*_{tot,x=L}. \tag{13a,b,c}$$

In Eq. (13c), $j^*_{tot,x=L}$ is the flux of tau into the terminal. Following ref. [31] (the details are given in section S2 in the Supplemental Materials), Eq. (13c) was restated as:

$$\text{At } x^* = L^*: \quad -D^*_{free}\frac{\partial n^*_{free}}{\partial x^*} - D^*_{mt}\frac{\partial n^*_{dif}}{\partial x^*} + v^*_a n^*_a - v^*_r n^*_r = A\left(1 - \exp\left[-\frac{\ln(2)}{T^*_{1/2}}\frac{1}{\gamma^*_{ar}}\right]\right)v^*_a n^*_a. \tag{14}$$

The amount of published experimental data for the axon exceeds the amount of published experimental data for the AIS. We therefore need to obtain as much information as possible before solving the problem in the AIS. We first solve the inverse problem in the axon to get data at the AIS/axon interface, which enables us to move on to solving the problem in the AIS.

Parameters that we were able to find based on data reported in literature are summarized in Table S1. Before the model given by Eqs. (1)-(14) can be used, the values of other parameters involved in the model need to be determined. This is an inverse problem [47,48]. Ref. [49] reported concentrations for two representative neurons, one with a longer (~600 μm) and the other with a shorter (~368 μm) axon. We used data for the longer axon (hollow circles in Fig. 3a) to determine values of model constants for the axon (fourth column in Table S2), and then we used data for a shorter axon (triangles in Fig. 3a) to validate the results. The dimensionless total tau concentration displayed in Fig. 3a is defined in Table S3.

In our problem, we use different types of published data, such as the tau concentration and velocity. Physically, tau concentrations in various kinetic states must also be non-negative. In order to solve the inverse problem we used multi-objective optimization (see, for example, [50-54]). In the axon (region 2) we estimated model parameters by finding the set of parameters that minimizes the following weighted objective (penalty) function which combines four different effects. We dropped ($i$=2) to simplify notation:

$$err2 = \sum_{j=1}^{N_2}\left(n_{tot,j} - n_{tot,exper,j}\right)^2 + \omega_1 \sum_{j=1}^{N_2}\left(v^*_{av,j} - 0.00345\frac{\mu\text{m}}{\text{s}}\right)^2 + \omega_2\left(\%\text{bound}\big|_{x=L/2} - 100\right)^2 + \omega_3, \quad (15)$$

where $N_2$=55, which is the number of data points that we obtained by scanning tau concentration data reported in Fig. 7D of [49]. In the second term on the right-hand side 0.00345 μm/s is the average tau velocity (which varies between 0.2-0.4 mm/day, motors-driven tau is transported in slow component a) reported in [55], also see [28]. Ref. [56] reported that ~99% of tau is bound to MTs. To account for this report, the third term on the right-hand side of Eq. (15) is used to ensure that in the center of the axon most tau is bound to MTs. We used $\omega_1$=10,000 s$^2$/μm$^2$ and $\omega_2$=1 so that the concentration predicted by the model is in good visual agreement with experimental data (hollow circles in Fig. 3a; also, a grey band in Fig. 4a). We also used $\omega_3 = 10^8$ if any of $n_{a,j}$, $n_{r,j}$, $n_{a0,j}$, $n_{r0,j}$, $n_{free,j}$, $n_{dif,j}$, and $n_{st,j}$ ($j$=1,…,55) are negative, which was done to prevent negative tau concentrations in any of the kinetic states anywhere in the axon. Values of these weighting factors were found by extensive numerical experimentation (data not shown). For example, a very large value of $\omega_1$ would cause an overfit in terms of tau average velocity, which would be forced to approach exactly 0.00345 μm/s. This would ignore the natural variance of the velocity [57]. Also, setting $\omega_1$ to a very large value would put more weight on the tau velocity at the cost of tau concentration, and would lead to predicting a total tau concentration that would fit the experimentally measured tau concentration reported in [49] worse.

For region 1 ($-L^*_{AIS} \le x^* \le 0$) Eqs. (1)-(7) were solved subject to the following boundary conditions:

At $x^* = -L^*_{AIS}$: $\quad j^*_{tot}\big|_{x=-L_{AIS}} = j^*_{tot,x=-L_{AIS}}, \quad n^*_{tot}\big|_{x=-L_{AIS}} = n^*_{tot,x=-L_{AIS}} = n^*_{tot,3}$. (16)

According to the data reported in Fig. 7D of [49], the minimum total concentration of tau in the axon occurs in the 3rd point, $n^*_{tot,3}$. We assumed that in the AIS the total tau concentration is changing between $n^*_{tot,3}$ and the concentration at the interface between the AIS and the axon.

At the interface between the AIS and the axon, we imposed the following boundary conditions:

At $x^*=0$: $\left.n^*_a\right|_{x=0^-}=\left.n^*_a\right|_{x=0^+}$, $\left.n^*_r\right|_{x=0^-}=\left.n^*_r\right|_{x=0^+}$, $\left.n^*_{tot}\right|_{x=0^-}=\left.n^*_{tot}\right|_{x=0^+}$,

$$\left.j^*_{tot}\right|_{x=0^-}=\left.j^*_{tot}\right|_{x=0^+}. \quad (17a,b,c)$$

The tau concentration in the axon is approximately two times greater than in the soma. Since the tau concentration is low in the soma and high in the axon, we assumed that the tau concentration increases from the beginning of the AIS to its end. Since an increased concentration of AnkG indicates the location of the AIS, by comparing the data showing tau distributions with the data showing AnkG distributions in Fig. 2.2 of [58] we deduced that the increase of the tau concentration in the AIS is approximately linear. The same linear increase of the tau concentration was deduced from Fig. 3(f) of [59]. We thus assumed a linear increase of the total tau concentration in the AIS (see crosses in Fig. 3a).

We estimated model parameters for the AIS (region 1) by minimizing the following weighted objective function. We dropped ($i$=1) to simplify notations:

$$\begin{aligned} err1 &= \sum_{j=1}^{N_1}\left(n_{tot,j}-n_{tot,exper,j}\right)^2+\omega_3+10^2\sum_{j=1}^{N_1}\left(n_{r,j}\right)^2 \\ &+10^5\left(n_{a,1}\right)^2+10^2\left(n_{r,1}\right)^2 \\ &+10\left(n_{a0,5}-n_{a0,x=0}\right)^2+10\left(n_{r0,5}-n_{r0,x=0}\right)^2+10^5\left(n_{free,5}-n_{free,x=0}\right)^2+10\left(n_{dif,5}-n_{dif,x=0}\right)^2+10\left(n_{st,5}-n_{st,x=0}\right)^2, \end{aligned} \quad (18)$$

where $N_1$=5. As with Eq. (15), values of weighting factors in Eq. (18) were found by extensive numerical experimentation (data not shown). The last term in the first line of Eq. (18) was used to minimize tau concentration in the kinetic state that simulates active retrograde transport. The terms in the second line of Eq. (18) were used to minimize the deviation of $n_a$ and $n_r$ in the beginning of the AIS from zero. The terms in the third line of Eq. (18) were used to minimize the deviation of values of $n_{a0}$, $n_{r0}$, $n_{free}$, $n_{dif}$, and $n_{st}$ at the interface between the AIS and the axon. For example, $n_{a0,5}$ denotes a value of $n_{a0}$ at the AIS side of the interface and $n_{a0,x=0}$ denotes the same value at the axon side of the interface.

Since we first solved the problem given by Eqs. (1)-(7) with boundary conditions (12)-(14) in the axon and minimized the objective function given by Eq. (15), we could use the solution obtained for the axon to determine values of $n_{a,x=0}$, $n_{r,x=0}$, $n_{tot,x=0}$, and $j^*_{tot,x=0}$ that we will need for imposing the boundary conditions at the AIS/interface, see Eq. (17). We also used the solution in the axon to determine values of $n_{a0,x=0}$, $n_{r0,x=0}$, $n_{free,x=0}$, $n_{dif,x=0}$, and $n_{st,x=0}$ that we need for Eq. (18). We then solved the problem given by Eqs. (1)-(7) with boundary conditions (16) and (17) in the AIS and minimized the objective function given by Eq. (18).

## 2.2. Numerical solution

Eqs. (1)-(7) are a mixture of ordinary differential equations, which contain various derivatives with respect to $x^*$, and algebraic equations. Algebraic Eqs. (3), (4), and (7) were used to eliminate $n^*_{a0}\left(x^*\right)$, $n^*_{r0}\left(x^*\right)$, and $n^*_{st}\left(x^*\right)$ from the remaining equations. The boundary value problem for the remaining equations, together with boundary conditions (12), (13) for the axon, and together with boundary conditions (16), (17) for the AIS, was solved by utilizing the Matlab's BVP4C solver (Matlab R2019a, MathWorks, Natick, MA, USA) [32]. We found this method to be more robust than solving a boundary value problem for a mixture of differential and algebraic equations directly.

In order to minimize the objective function given by Eq. (15), we used MULTISTART with the local solver FMINCON; these routines are included in Matlab R2019a, MathWorks, Natick, MA, USA's Optimization Toolbox (Matlab R2019a, MathWorks, Natick, MA, USA). The default interior-point algorithm option in FMINCON was used [60]. 10,000 randomly selected starting points were used in MULTISTART. The same was done for Eq. (18).

# 3. Results

## 3.1. Investigating tau concentrations and fluxes in the axon and the AIS

All tau concentrations were non-dimensionalized as given in Table S3. Due to a small degradation rate of tau in the axon, the total flux of tau is almost independent of the position in the axon (Fig. 3b). The shape of the average velocity of tau (Fig. 4a) is explained by the fact that the tau velocity is defined as the ratio of tau flux to tau concentration (Eq. (11)).

In most of the axon (away from the AIS and from the terminal), the concentration of tau that is driven by molecular motors anterogradely stays at a constant value (Fig. 4b). This means that the anterograde flux of tau in most of the axon (away from the boundaries) remains constant and independent of the length of the axon (Fig. 5b). The same applies to the retrograde motor-driven flux of tau (Fig. S1a, see also Fig. 5b). This suggests that the values of the model constants, γs, reported in the fourth column of Table S2, are applicable to any axon independent of its length. The independence of γs from the length of the axon means that our model has some generality. It should be noted that the concentrations of anterograde and retrograde pausing tau, $n_{a0}$ and $n_{r0}$, also stay constant in most of the axon (Figs. S1b and S2a). This is because when $n_a$ and $n_r$ are independent of *x*, $n_{a0}$ is directly proportional to $n_a$, and $n_r$ is directly proportional to $n_{r0}$ (Eqs. (1) and (2)).

Intriguingly, the concentration of free (cytosolic) tau in the AIS decreases with an increase in distance from the soma (Fig. 5a). This result is counter-intuitive since the tau concentration in the axon is about two times greater than in the soma ([58], see also Fig. 3a). The decrease of the free tau concentration in the AIS in Fig. 5a is explained by a greater affinity of tau to MTs down the AIS. As more tau attaches to MTs, less free tau remains (Fig. S3b). For example, the concentration of stationary tau bound to MTs increases in the AIS as the distance from the soma increases (Fig. S2b). Also, the concentration of MT-bound tau that can diffuse along the MTs (with a ~70 times less diffusivity than diffusivity of free tau, Table S1) increases in the AIS as the distance from the soma increases (Fig. 3a).

The negative gradient of free tau in the AIS (Fig. 5a) results in a diffusion-driven transport of free tau from the soma into the axon. This important prediction of the model should be validated through future experiments.

The diffusion-driven flux of free tau is the largest at the boundary between the soma and the AIS. As the distance from the soma increases, the diffusion-driven flux of tau decreases and the flux driven by anterograde motors increases (Fig. 5b). This means that tau uses the diffusion-driven mode of transport to enter the AIS, but once it is in the AIS, it gradually switches to the motor-driven mode.

### 3.2. Investigating sensitivity of the DCV flux into AIS on the kinetic constants

We investigated how the flux into the AIS depends on the model parameters, γs, by computing the local sensitivity coefficients, which are first-order partial derivatives of the flux with respect to γs [47,61-63].

For example, the sensitivity coefficient of $j_{tot,x=-L_{AIS}}$ to parameter $\gamma^{*}_{off,a(1)}$ was calculated as follows:

$$\frac{\partial j_{tot,x=-L_{AIS}}}{\partial \gamma^*_{off,a(1)}} \approx \left. \frac{j_{tot,x=-L_{AIS}}\left(\gamma^*_{off,a(1)} + \Delta\gamma^*_{off,a(1)}\right) - j_{tot,x=-L_{AIS}}\left(\gamma^*_{off,a(1)}\right)}{\Delta\gamma^*_{off,a(1)}} \right|_{\text{other parmeters kept constant}}, \qquad (19)$$

where $\Delta\gamma^*_{off,a(1)} = 10^{-3}\gamma^*_{off,a(1)}$ (the independence of $\Delta\gamma^*_{off,a(1)}$ was tested by using various step sizes).

Non-dimensionalized relative sensitivity coefficients were calculated following [61,64] as (for example):

$$S^{j_{tot,x=-L_{AIS}}}_{\gamma^*_{off,a(1)}} = \frac{\gamma^*_{off,a(1)}}{j_{tot,x=-L_{AIS}}} \frac{\partial j_{tot,x=-L_{AIS}}}{\partial \gamma^*_{off,a(1)}}. \qquad (20)$$

Parameters involved in Eqs. (19) and (20) are defined in Table S2. The sensitivity coefficients are reported in Table 1.

Table 1. Dimensionless sensitivity coefficients for the DCV flux from the soma into the AIS, $j_{tot,x=-L_{AIS}}$. Computations were performed with $\Delta\gamma^* = 10^{-3}\gamma^*$. An almost identical result was obtained for $\Delta\gamma^* = 10^{-2}\gamma^*$. The cells with the largest sensitivity coefficients are shaded grey.

| $S^{j_{tot,x=-L_{AIS}}}_{\gamma^*_{off,a(1)}}$ | $S^{j_{tot,x=-L_{AIS}}}_{\gamma^*_{off,r(1)}}$ | $S^{j_{tot,x=-L_{AIS}}}_{\gamma^*_{ra(1)}}$ | $S^{j_{tot,x=-L_{AIS}}}_{\gamma^*_{ar(1)}}$ | $S^{j_{tot,x=-L_{AIS}}}_{\gamma^*_{dif->st(1)}}$ | $S^{j_{tot,x=-L_{AIS}}}_{\gamma^*_{st->dif(1)}}$ | $S^{j_{tot,x=-L_{AIS}}}_{\gamma^*_{10(1)}}$ |
|---|---|---|---|---|---|---|
| $-4.575\times10^{-4}$ | $-2.6962\times10^{-5}$ | $3.053\times10^{-2}$ | $-1.226\times10^{-3}$ | $1.935\times10^{-2}$ | $-1.883\times10^{-2}$ | $2.124\times10^{-2}$ |

| $S^{j_{tot,x=-L_{AIS}}}_{\gamma^*_{01(1)}}$ | $S^{j_{tot,x=-L_{AIS}}}_{\gamma^*_{on,a(1)}}$ | $S^{j_{tot,x=-L_{AIS}}}_{\gamma^*_{on,r(1)}}$ | $S^{j_{tot,x=-L_{AIS}}}_{\gamma^*_{free->st(1)}}$ | $S^{j_{tot,x=-L_{AIS}}}_{\gamma^*_{st->free(1)}}$ | $S^{j_{tot,x=-L_{AIS}}}_{\gamma^*_{dif->free(1)}}$ | $S^{j_{tot,x=-L_{AIS}}}_{\gamma^*_{free->dif(1)}}$ |
|---|---|---|---|---|---|---|
| $-2.882\times10^{-2}$ | $4.858\times10^{-1}$ | $2.710\times10^{-3}$ | $-1.460\times10^{-2}$ | $1.883\times10^{-2}$ | $2.929\times10^{-3}$ | $-1.150\times10^{-2}$ |

The results reported in Table 1 show that the total flux of tau is most sensitive to parameter $\gamma_{on,a(1)}$, which describes the rate of tau transition from the free (cytosolic) to anterograde pausing state. The sensitivity coefficient $S^{j_{tot,x=-L_{AIS}}}_{\gamma^*_{ra(1)}}$ is positive. The reduction of $\gamma_{on,a(1)}$ by a factor of two reduces the total flux of tau in the AIS by 26.7% (Fig. S4a). It also reduces the average velocity of tau in the AIS (Fig. S4b). This happens because the amount of tau moved anterogradely by molecular motors is reduced (Fig. S5a), but the amount of tau in the free (cytosolic) state, where transport is less efficient, is increased (Fig. S5b).

The other parameters to which $j_{tot}$ exhibits a large positive sensitivity ( $j_{tot}$ increases when the parameter increases) are $\gamma_{ra(1)}$, $\gamma_{dif\to st(1)}$, $\gamma_{10(1)}$, and $\gamma_{st\to free(1)}$. The other parameters to which $j_{tot}$ exhibits a large negative sensitivity ( $j_{tot}$ decreases when the parameter increases) are $\gamma_{st\to dif(1)}$, $\gamma_{01(1)}$, and $\gamma_{free\to st(1)}$ (Table 1).

**4. Discussion and future directions**

By fitting the model with experimental data in the axon, and by assuming a linear tau distribution in the AIS, we determined kinetic constants for the model of tau transport in the axon and in the AIS. This model was used to obtain the following predictions concerning tau transport.

We found that in most of the axon the fluxes of tau driven by anterograde and retrograde motors are independent of the distance from the soma. This means that the values of kinetic constants determined in our research can be applied to an axon of any length.

We also found that the concentration of free (cytosolic) tau decreases in the AIS as the distance from the soma increases. This is because with an increase in distance from the soma, more monomers of free tau bind to MTs. The negative gradient of free tau results in diffusion-driven flux of free tau into the AIS.

Our model thus explains how tau protein can move from the somatodendritic compartment to the axonal compartment [28]. The positive gradient of the total tau concentration in the AIS (Fig. 3a) does not result in tau diffusion to the soma. This is because binding of free tau to MTs creates a negative gradient of free tau in the AIS (Fig. 5a), which results in diffusion-driven tau transport from the soma into the AIS (Fig. 5b).

Our modeling results suggest that the largest diffusion-driven flux of tau occurs at the boundary between the soma and the AIS ( $x^* = -L^*_{AIS}$ ). The motor-driven anterograde flux of tau in this location is very small. As the distance from the soma increases, the diffusion-driven flux of free tau decreases and the anterograde motor-driven flux increases, which suggests that tau switches from a diffusion-driven to the motor-driven mode of transport. This is consistent with [41], which suggested that slow axonal transport mechanism dominates at longer distances.

Our modeling results clarify how tau transport in the AIS may function. It was previously suggested that diffusion of free tau out of the AIS is counteracted by larger motor-driven anterograde transport of tau into the AIS, see the "Directional transport model" in Fig. 9A of [27]. However, [65] pointed out that the inefficiency of slow axonal transport in moving tau from the soma to the axon makes it difficult to explain how it can work against diffusion transport of tau, which is very efficient because of tau's large diffusivity. Our results suggest that motor-driven and diffusion-driven transport work together in the AIS. Both mechanisms transport tau anterogradely, and there is an interplay between them: as the distance from the soma increases, the diffusion-driven transport becomes smaller, while motor-driven transport increases. The AIS thus works not only as a barrier, but also as a pump that moves tau into the axon. For that reason, we believe that TDB would be better called a tau retrograde barrier.

This is an important insight because loss of tau enrichment in the axon, which is likely caused by malfunction of the TDB, is an early sign of AD [24,66]. Since tau concentration in the axon is larger than in the soma, without the TDB tau diffusion will move tau from the axon to the soma, which is exactly what happens at the onset of AD. The TDB is expected to prevent retrograde diffusion of tau in some way or overpower it by larger motor-driven anterograde transport [27]. Our results suggest that in a healthy AIS diffusion and motor-driven transport can both transport tau anterogradely. The model also predicts a potential interplay between these two transport mechanisms (when the diffusion-driven flux of tau becomes smaller, the motor-driven flux becomes larger). Understanding normal operation of the TDB is important for understanding what can go wrong at the onset of AD.

Our model also suggests that the total tau flux entering the AIS is most sensitive to the kinetic constant that describes tau transition from the free to the anterograde pausing state. A reduction of this constant reduces the tau flux because tau transport by anterograde motors (which is most efficient) is reduced.

Future experimental research is needed to check theoretical predictions reported in this paper. Also, in a human brain, tau occurs in six isoforms. The tightness of the tau retrograde barrier depends on the isoform. Longer tau isoforms cannot pass the barrier in the retrograde direction while shorter isoforms can, at least to a degree [67]. Future research should extend our model to be isoform-specific.

**Acknowledgment**

IAK acknowledges the fellowship support of the Paul and Daisy Soros Fellowship for New Americans and the NIH/National Institutes of Mental Health (NIMH) Ruth L. Kirchstein NRSA (F30 MH122076-01). AVK acknowledges the support of the National Science Foundation (award CBET-1642262) and the Alexander von Humboldt Foundation through the Humboldt Research Award.

**Figure captions**

Fig. 1. A diagram of a neuron showing various modes of slow axonal transport and diffusion of tau protein. The figure shows two regions used in the mathematical model: the AIS (region 1) and the axon (region 2). In displaying possible mechanisms of tau transport, we followed ref. [28] (see Fig. 3 in [28]). Tau can be transported by molecular motors (active transport), or by a diffusion-driven mechanism, either in the cytosol or by diffusion along MTs.

Fig. 2. A kinetic diagram showing various kinetic states in our model of tau transport in the axon and in the AIS.

Fig. 3. (a) Total concentration of tau. Experimental data from Fig. 7D of [49] are shown by open circles. These data were rescaled such that the experimentally measured tau concentration at the boundary between the AIS and the axon (at $x$=0) was equal to 1. Crosses show interpolated data across the AIS. Triangles show data for a shorter axon, given in Fig. 7B of [49], which were not used for determining model constants. (b) Total flux of tau, versus position in the axon. The vertical green line at $x$=0 shows the boundary between the AIS and the rest of the axon.

Fig. 4. (a) Average velocity of tau. The range of the average tau velocity reported in [55] is shown by a horizontal band. (b) Concentration of on-track tau transported by molecular motors in the anterograde direction, versus position in the axon. The vertical green line at $x$=0 shows the boundary between the AIS and the rest of the axon.

Fig. 5. (a) Concentration of free (diffusing) tau. (b) Various components of the tau flux: the motor-driven anterograde component; the motor-driven retrograde component; the component due to diffusion of cytosolic tau, $-D_{free}\frac{dn_{free}}{dx}$; and the component due to diffusion of MT-bound tau, $-D_{mt}\frac{dn_{mt}}{dx}$; versus position in the axon. The vertical green line at $x$=0 shows the boundary between the AIS and the rest of the axon.

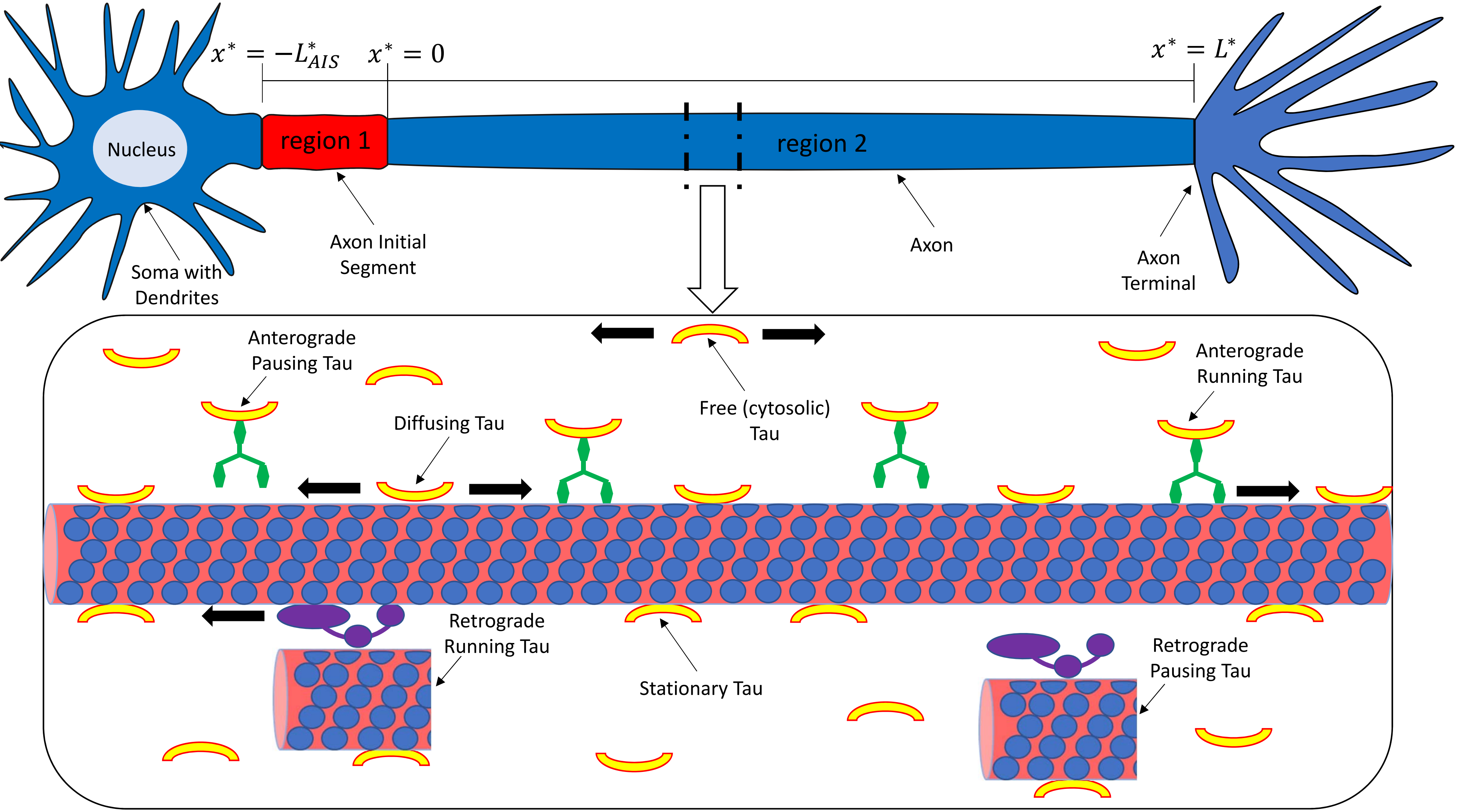

Figure 1

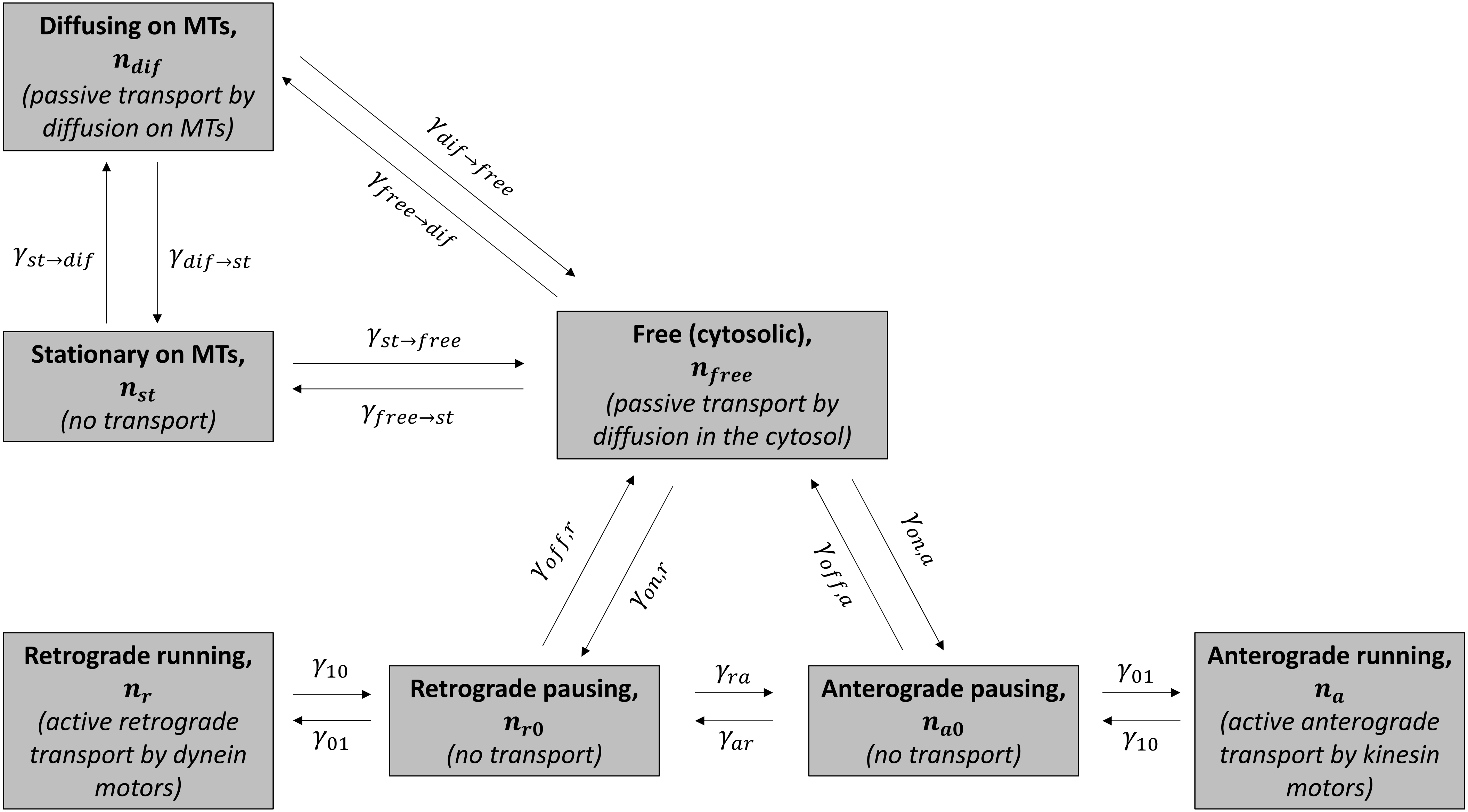

Figure 2

(a)

(b)

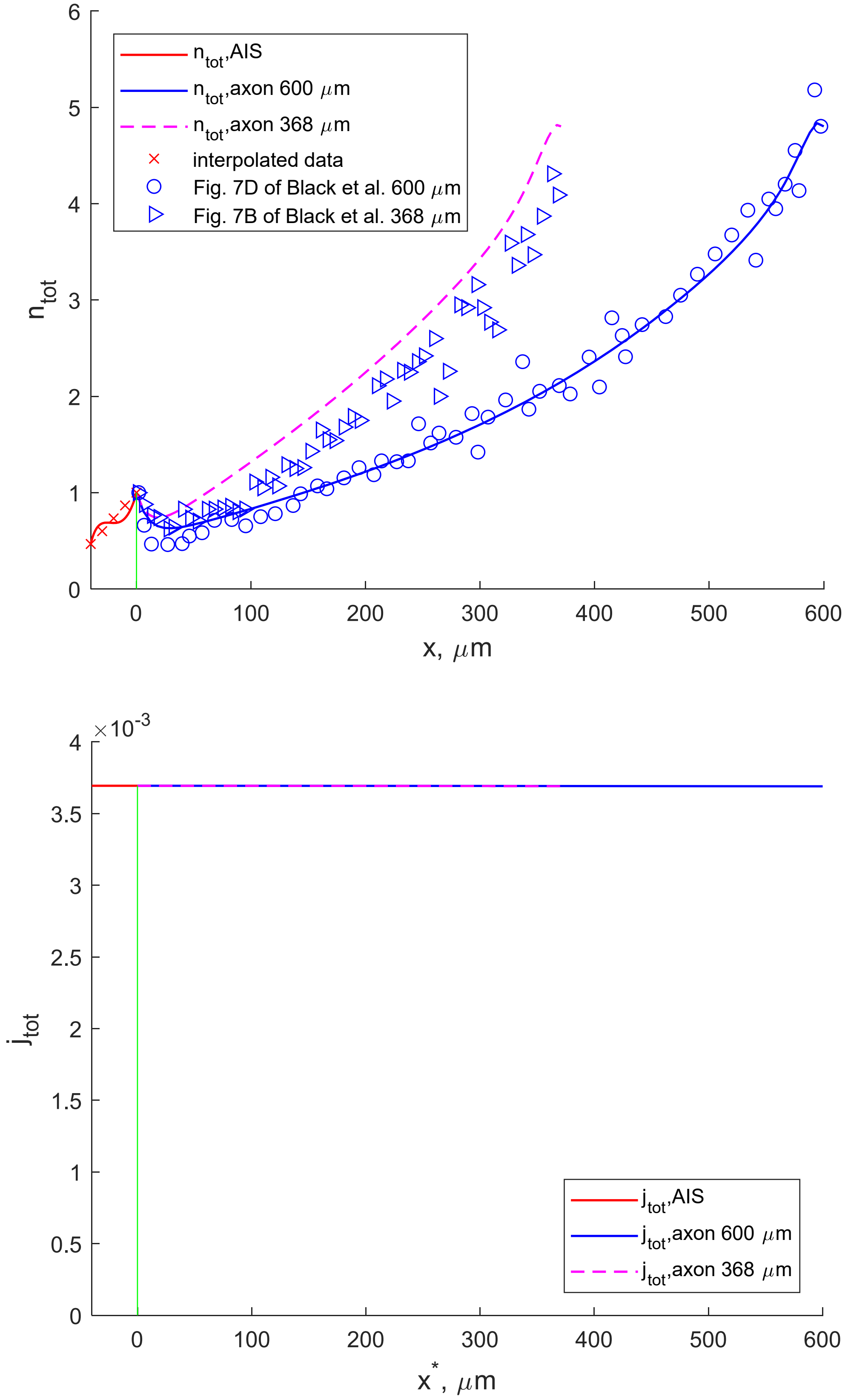

6
5
4
3
2
1
0
n_tot
n_tot,AIS
n_tot,axon 600 μm
n_tot,axon 368 μm
interpolated data
Fig. 7D of Black et al. 600 μm
Fig. 7B of Black et al. 368 μm
0
100
200
300
400
500
600
x, μm
×10^-3
4
3.5
3
2.5
2
1.5
1
0.5
0
j_tot
j_tot,AIS
j_tot,axon 600 μm
j_tot,axon 368 μm
x*, μm

Figure 3

(a)

(b)

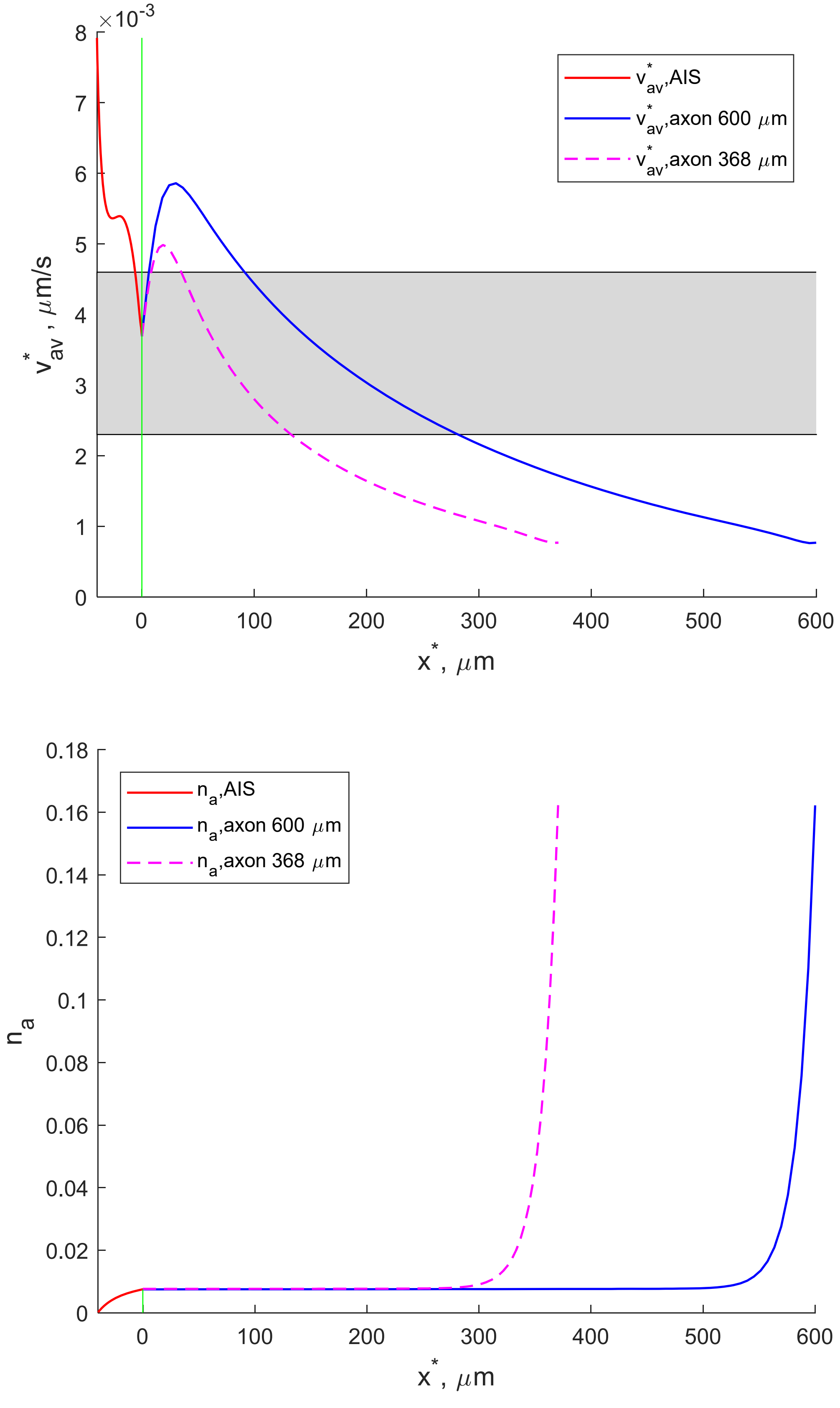

Figure 4

(a)

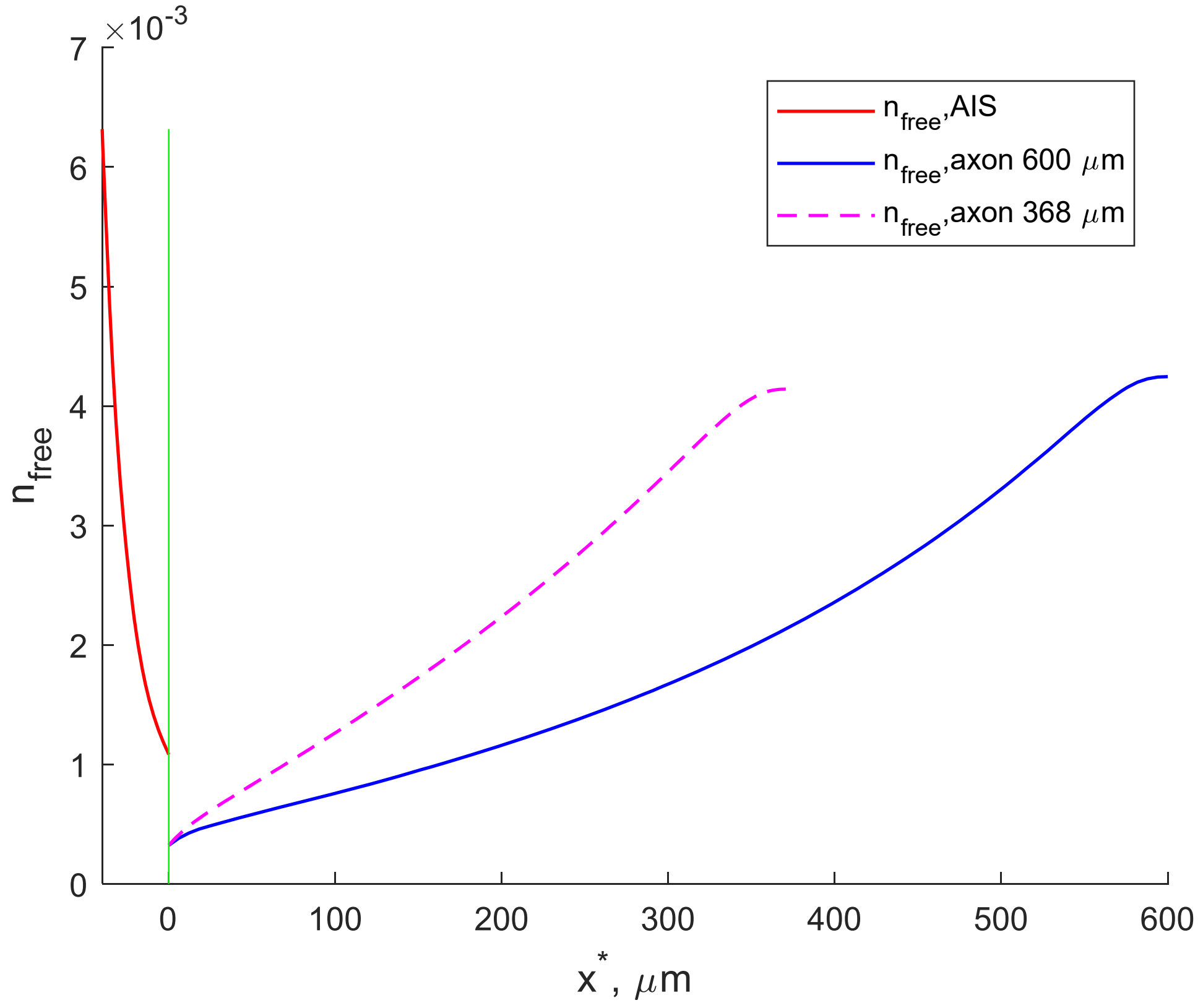

(b)

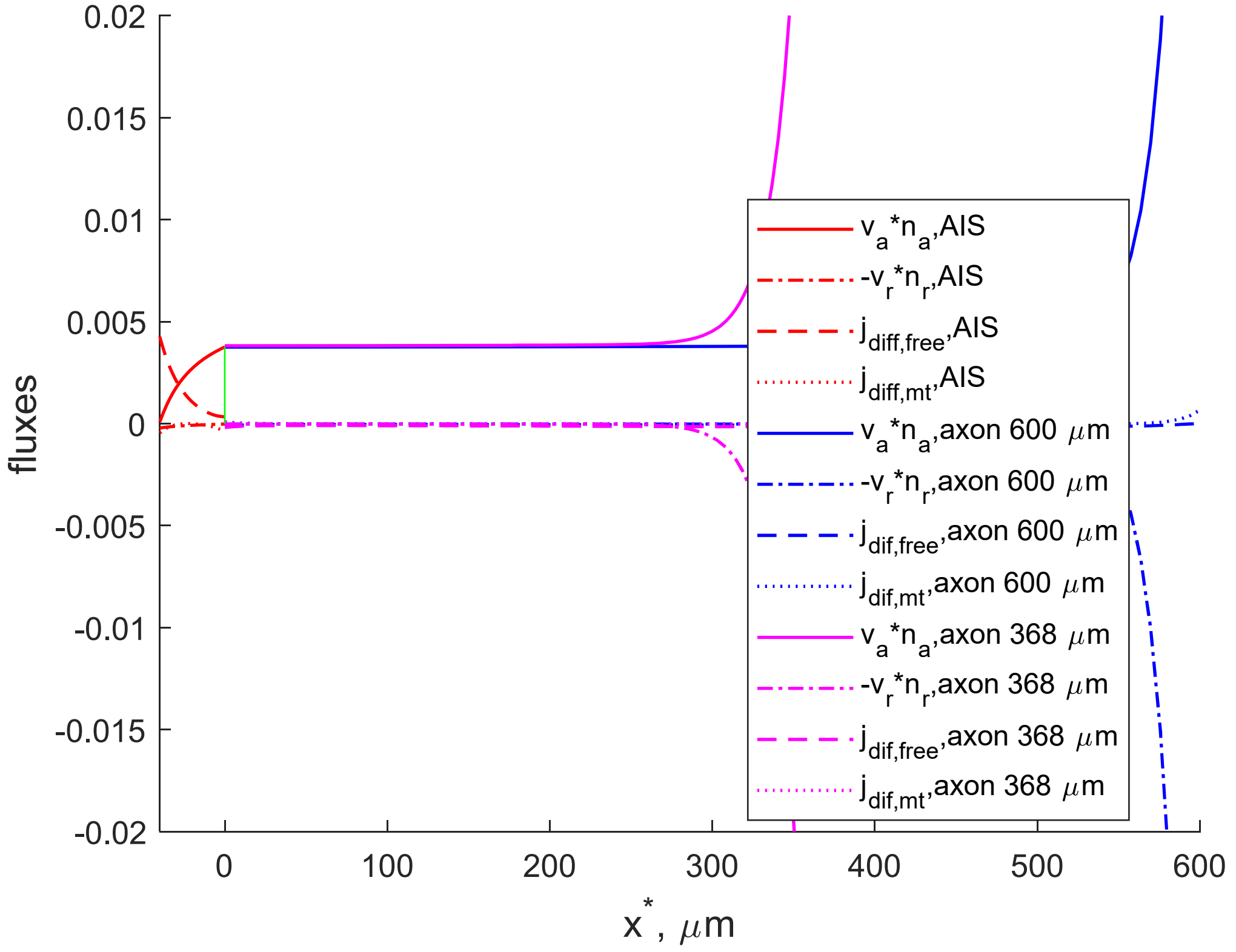

Figure 5

**Modeling tau transport in the axon initial segment**

I. A. Kuznetsov[(a), (b)] and A. V. Kuznetsov[(c)]

[(a)]Perelman School of Medicine, University of Pennsylvania, Philadelphia, PA 19104, USA

[(b)]Department of Bioengineering, University of Pennsylvania, Philadelphia, PA 19104, USA

[(c)]Dept. of Mechanical and Aerospace Engineering, North Carolina State University, Raleigh, NC 27695-7910, USA; e-mail: avkuznet@ncsu.edu

## Supplementary material

### S1. Supplementary tables

Table S1. Parameters of the model that we estimated based on values found in literature.

| Symbol | Definition | Units | Estimated value | Reference or estimation method |
|---|---|---|---|---|
| $D^*_{free}$ | Diffusivity of tau protein in the cytosolic state | $\mu m^2/s$ | 11[a] | [24,39] |
| $D^*_{mt}$ | Diffusivity of tau protein along MTs | $\mu m^2/s$ | 0.153[b] | [41] |
| $L^*$ | Length of the axon (without the AIS) | μm | 368; 600 | [44] |
| $L^*_{AIS}$ | Length of the AIS | μm | 40 | [1,6] |
| $T^*_{1/2}$ | Half-life of free monomeric tau protein | s | $2.16\times10^5$ | [68] |
| $v^*_a$, $v^*_r$ | Velocities of rapid motions of tau on MTs propelled by | μm/s | 0.5, 0.5 | [39] |
| | kinesin and dynein motors, respectively | | | |

[a] Pseudo-phosphorylated tau can diffuse with a diffusion coefficient of 11 μm$^2$/s, although the apparent diffusivity of tau is 3 μm$^2$/s [24,39]. The reduction of the apparent tau diffusivity is due to binding of tau to MTs [24], which leads to reduction of the portion of free tau.

[b] Ref. [41] reported that in addition to free diffusion of tau, tau can also diffuse along MTs, with the average diffusion coefficient of 0.153 μm$^2$/s. Approximately half of tau monomers can diffuse bidirectionally along MTs, while the rest remain stationary. Transitions between stationary and mobile pools of tau occur rarely with a rate of 0.015 s$^{-1}$. The existence of mobile and stationary states for MT-bound tau is explained by the existence of different binding confirmations and different binding sites for tau.

Table S2. Parameters characterizing transport of tau protein in the axon (region 2) and in the AIS (region 1) and their estimated values. The values reported here are different from those reported in [31] because we now used a different value of $D^*_{free}$ and different boundary conditions. LSR (Least Square Regression) [48].

| Symbol | Definition | Units | Estimated value for the axon (region 2) | Estimated value for the AIS (region 1) | Reference or estimation method |
|---|---|---|---|---|---|
| $\gamma^*_{10}$ | Kinetic constant describing the rate of transitions $n^*_a \rightarrow n^*_{a0}$ and $n^*_r \rightarrow n^*_{r0}$ | $s^{-1}$ | $4.877\times10^{-2}$ [a] | $7.189\times10^{-2}$ [a] | LSR |
| $\gamma^*_{01}$ | Kinetic constant describing the rate of | $s^{-1}$ | $3.623\times10^{-3}$ [a] | $5.964\times10^{-3}$ [a] | LSR |
| | transitions $n_{a0}^* \rightarrow n_a^*$ and $n_{r0}^* \rightarrow n_r^*$ | | | | |
| $\gamma_{ar}^*$ | Kinetic constant describing the rate of transition $n_{a0}^* \rightarrow n_{r0}^*$ | $s^{-1}$ | $4.111\times10^{-5}$ | $4.033\times10^{-6}$ | LSR |
| $\gamma_{ra}^*$ | Kinetic constant describing the rate of transition $n_{r0}^* \rightarrow n_{a0}^*$ | $s^{-1}$ | $8.393\times10^{-3}$ | $1.898\times10^{-3}$ | LSR |
| $\gamma_{on,a}^*$ | Kinetic constant describing the rate of transition $n_{free}^* \rightarrow n_{a0}^*$ | $s^{-1}$ | $8.894\times10^{-5}$ | $3.277\times10^{-2}$ | LSR |
| $\gamma_{on,r}^*$ | Kinetic constant describing the rate of transition $n_{free}^* \rightarrow n_{r0}^*$ | $s^{-1}$ | $6.940\times10^{-5}$ | $3.908\times10^{-3}$ | LSR |
| $\gamma_{off,a}^*$ | Kinetic constant describing the rate of transition $n_{a0}^* \rightarrow n_{free}^*$ | $s^{-1}$ | $2.343\times10^{-8}$ | $1.786\times10^{-6}$ | LSR |
| $\gamma_{off,r}^*$ | Kinetic constant describing the rate of transition $n_{r0}^* \rightarrow n_{free}^*$ | $s^{-1}$ | $9.115\times10^{-5}$ | $3.272\times10^{-5}$ | LSR |
| $\gamma_{free\rightarrow st}^*$ | Kinetic constant describing the rate of transition $n_{free}^* \rightarrow n_{st}^*$ | $s^{-1}$ | $6.579\times10^{-3}$ | $9.166\times10^{-3}$ | LSR |
| $\gamma_{st\rightarrow free}^*$ | Kinetic constant describing the rate of transition $n_{st}^* \rightarrow n_{free}^*$ | $s^{-1}$ | $1.437\times10^{-6}$ | $5.821\times10^{-5}$ | LSR |
| $\gamma^*_{free\to dif}$ | Kinetic constant describing the rate of transition $n^*_{free} \to n^*_{dif}$ | $s^{-1}$ | $7.065\times10^{-3}$ | $6.450\times10^{-3}$ | LSR |
| $\gamma^*_{dif\to free}$ | Kinetic constant describing the rate of transition $n^*_{dif} \to n^*_{free}$ | $s^{-1}$ | $1.028\times10^{-3}$ | $1.458\times10^{-4}$ | LSR |
| $\gamma^*_{dif\to st}$ | Kinetic constant describing the rate of transition $n^*_{dif} \to n^*_{st}$ | $s^{-1}$ | $9.974\times10^{-3}$ | $8.988\times10^{-3}$ | LSR |
| $\gamma^*_{st\to dif}$ | Kinetic constant describing the rate of transition $n^*_{st} \to n^*_{dif}$ | $s^{-1}$ | $1.322\times10^{-4}$ | $4.851\times10^{-4}$ | LSR |
| $j_{tot,tau,x=0}$ [b] | Dimensionless total flux of tau from the AIS into the axon | | $3.694\times10^{-3}$ | | LSR |
| $n_{free,x=0}$ [c] | Dimensionless concentration of free (cytosolic) tau at the axon hillock | | $3.223\times10^{-4}$ | | LSR |
| $A$ | Coefficient in Eq. (14) | | $6.058\times10^{-1}$ | | LSR |
| $j_{tot,tau,x=-L_{AIS}}$ | Dimensionless total flux of tau from the soma into the AIS | | | $3.694\times10^{-3}$ | LSR |

[a] It should be noted that $\gamma^*_{10}$ and $\gamma^*_{01}$ are effective values of kinetic constants that give the best fit of model predictions with experimental or interpolated data. The dwell time of tau on MTs may be as short as 40 ms [56,69].

[b] $j_{tot,tau,x=0} = \dfrac{j^*_{tot,tau,x=0}}{n^*_{tot,ax,x=0} v^*_a}$.

[c] $n_{free,x=0} = \dfrac{n^*_{free,x=0}}{n^*_{tot,ax,x=0}}$.

Table S3. Dimensionless dependent variables in the model of tau transport.

| Symbol | Definition |
|---|---|
| $j_{tot,tau}$ | $\dfrac{j^*_{tot,tau}}{n^*_{tot,ax,x=0} v^*_a}$ |
| $n_{tot}$ | $\dfrac{n^*_{tot}}{n^*_{tot,ax,x=0}}$ |
| $n_a$ | $\dfrac{n^*_a}{n^*_{tot,ax,x=0}}$ |
| $n_r$ | $\dfrac{n^*_r}{n^*_{tot,ax,x=0}}$ |
| $n_{a0}$ | $\dfrac{n^*_{a0}}{n^*_{tot,ax,x=0}}$ |
| $n_{r0}$ | $\dfrac{n^*_{r0}}{n^*_{tot,ax,x=0}}$ |
| $n_{free}$ | $\dfrac{n^*_{free}}{n^*_{tot,ax,x=0}}$ |
| $n_{st}$ | $\dfrac{n^*_{st}}{n^*_{tot,ax,x=0}}$ |
| $n_{dif}$ | $\dfrac{n^*_{dif}}{n^*_{tot,ax,x=0}}$ |

## S2. Derivation of Eq. (14) by estimating the portion of tau degraded at the axon terminal

The approach developed in [70], which was developed for modeling transport of neurofilament proteins, was used here to recast the right-hand side of Eq. (13c) in a more detailed form. There are two possible fates for tau that enters the axon terminal: either it can be degraded or it can reverse its direction and re-enter the axon, moving retrogradely. The probability of tau degradation is estimated as $1-\exp\left[-\gamma_{deg}^{*} t_{rev}^{*}\right]$, where $\gamma_{deg}^{*}$ is a kinetic constant characterizing tau degradation rate and $t_{rev}^{*}$ is the time required for a motor-driven tau protein to reverse its direction in the terminal. The value of $\gamma_{deg}^{*}$ is estimated as $\ln(2)/T_{1/2}^{*}$ and the value of $t_{rev}^{*}$ is estimated as $1/\gamma_{ar(2)}^{*}$. We used the values of parameters $T_{1/2}^{*}$ and $\gamma_{ar(2)}^{*}$ that are given in Tables S1 and S2, and estimated the probability of tau degradation at the axon terminal to be 7.51%. This means that the remaining 92.49% of tau reverse its direction and leave the axon terminal, moving in the retrograde direction. Since this estimate is approximate, we multiplied this estimate by a parameter *A*, as we did in [71]. The value of parameter *A* is determined by finding the best fit with published experimental data. The procedure described above leads to Eq. (14).

## S3. Supplementary figures

### S3.1. Tau concentrations in various kinetic states for two different axon lengths

(a)

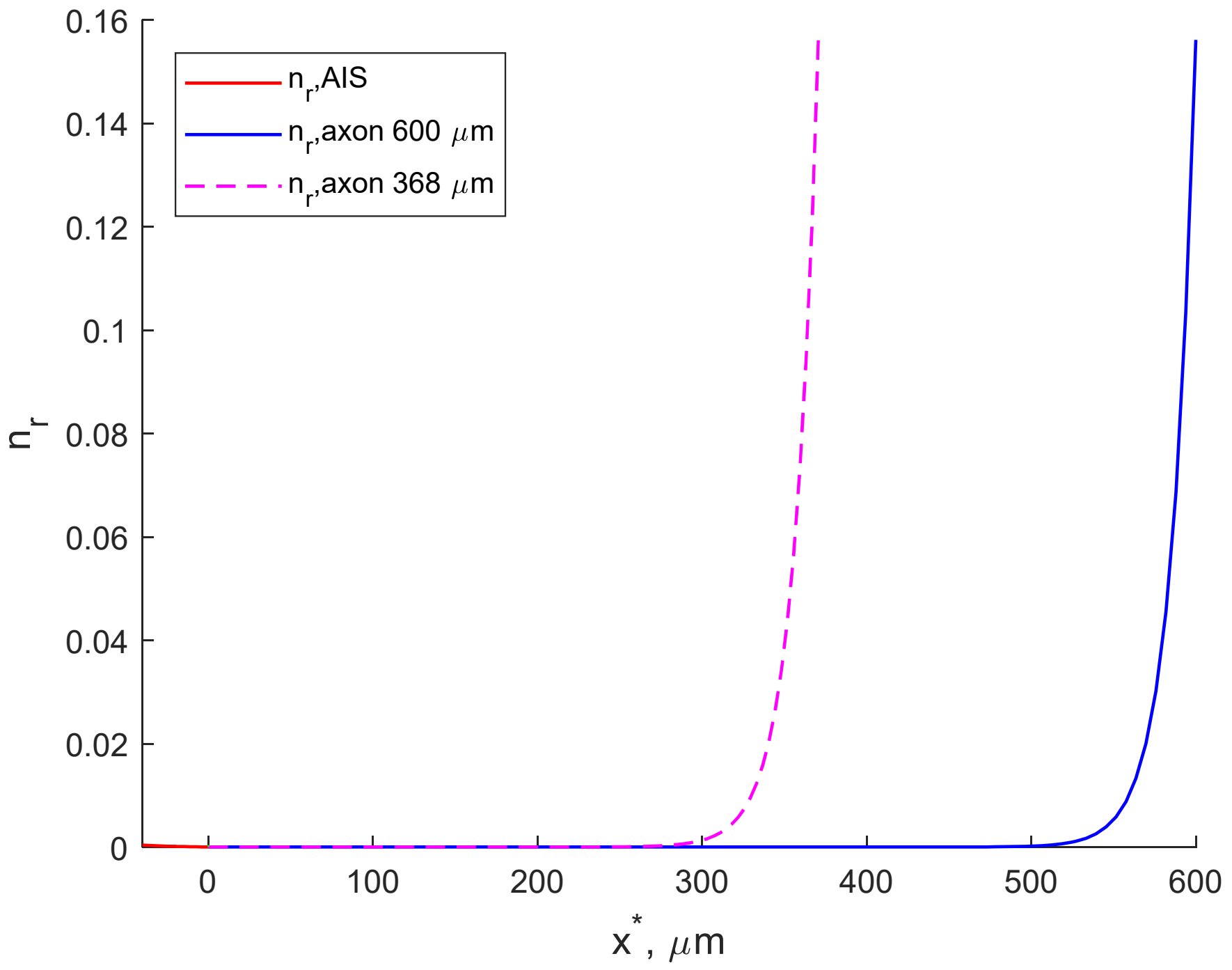

(b)

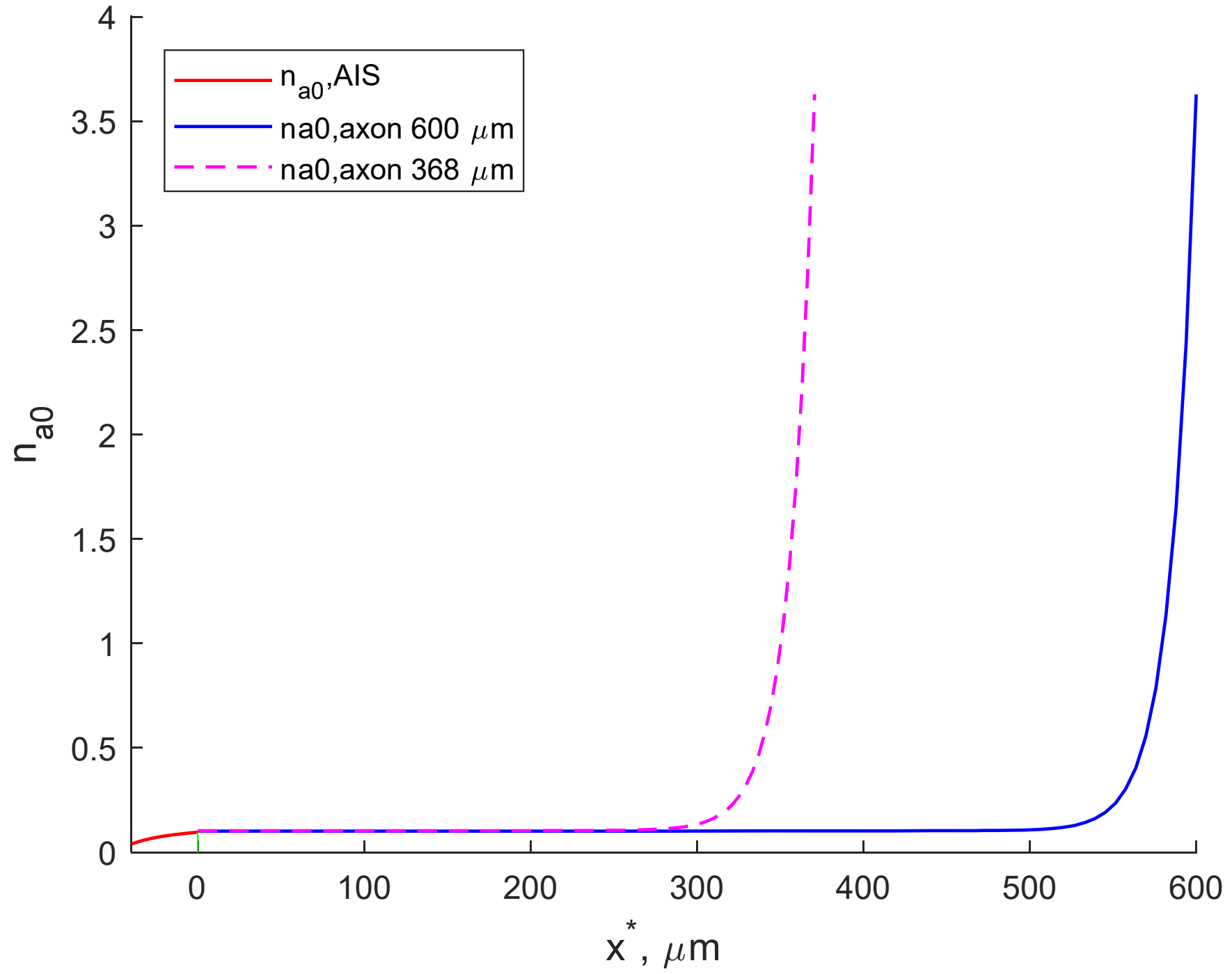

Fig. S1. (a) Concentration of on-track tau transported by molecular motors in the retrograde direction. (b) Concentration of pausing on-track tau associated with anterograde motors, versus position in the axon. The vertical green line at *x*=0 shows the boundary between the AIS and the rest of the axon.

(a)

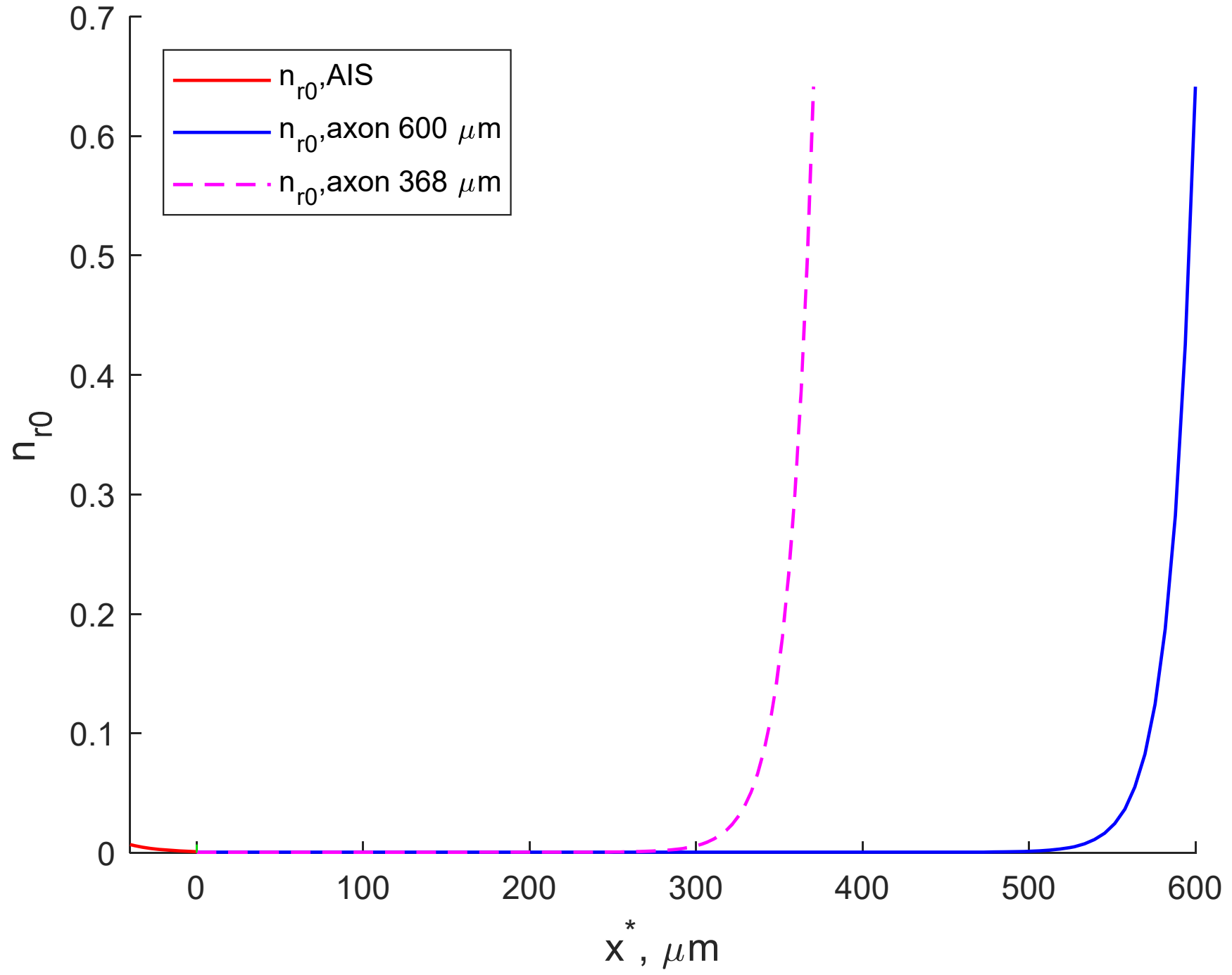

(b)

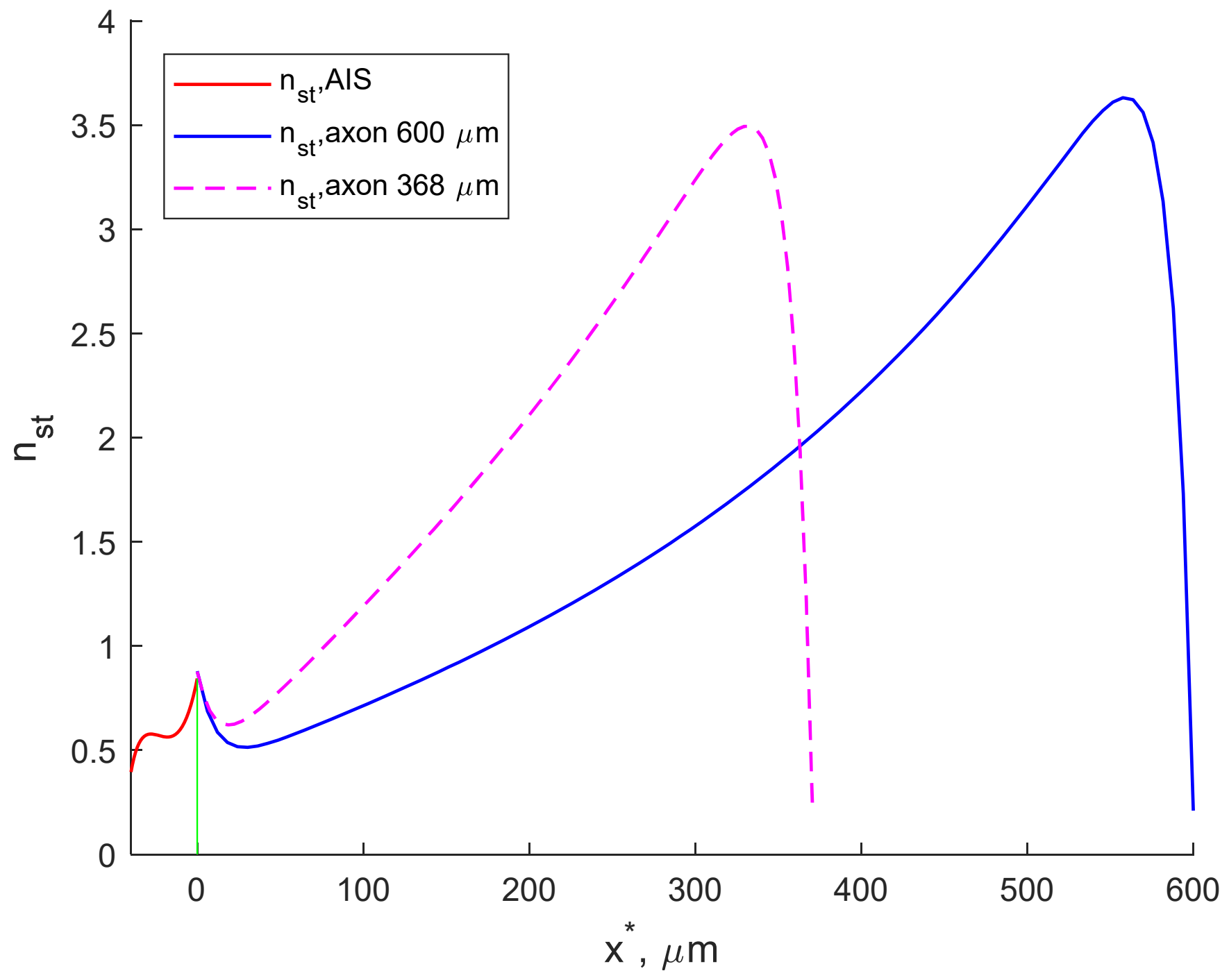

Fig. S2. (a) Concentration of pausing on-track tau associated with retrograde motors. (b) Concentration of stationary tau bound to MTs, no association with motors, versus position in the axon. The vertical green line at $x$=0 shows the boundary between the AIS and the rest of the axon.

(a)

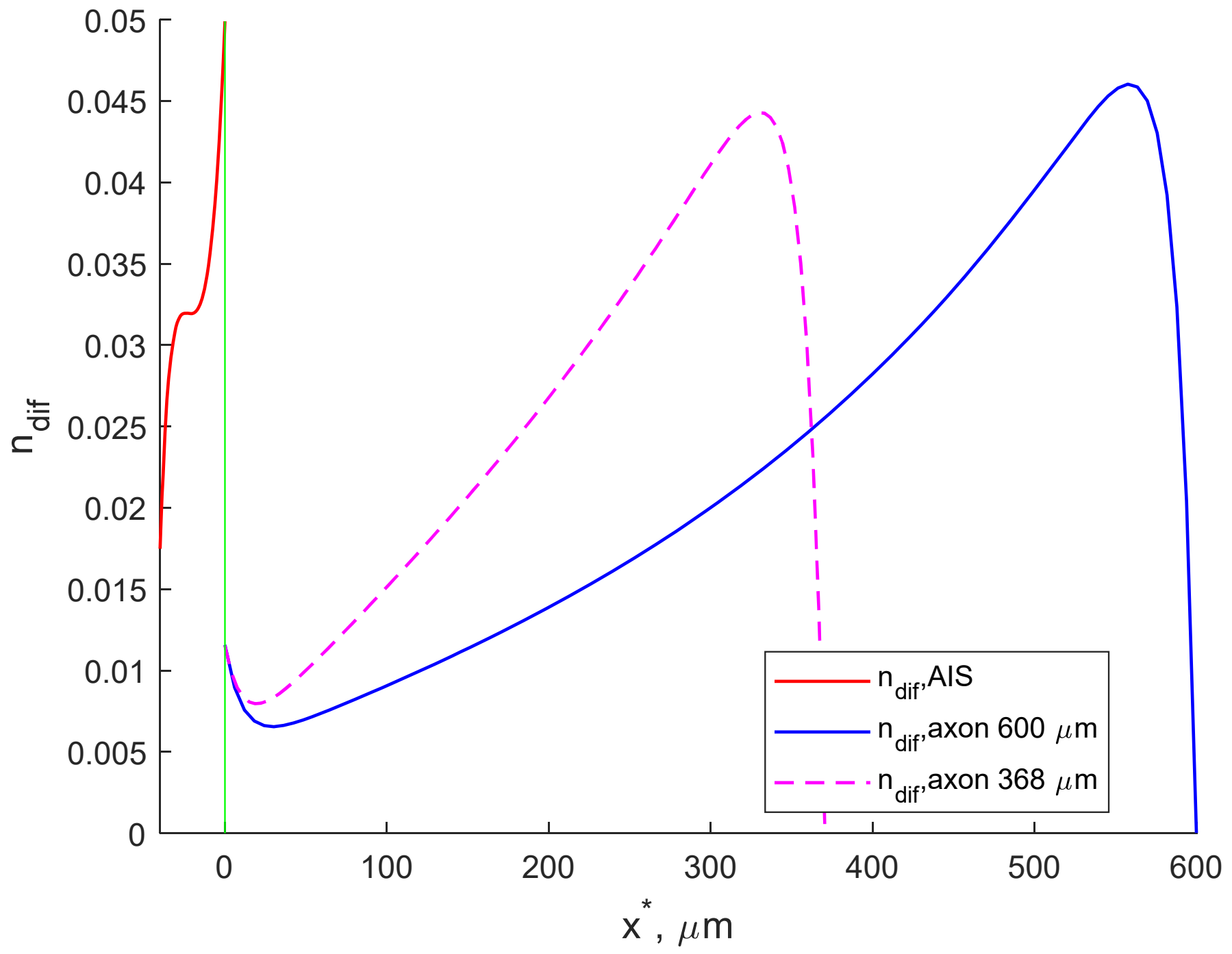

(b)

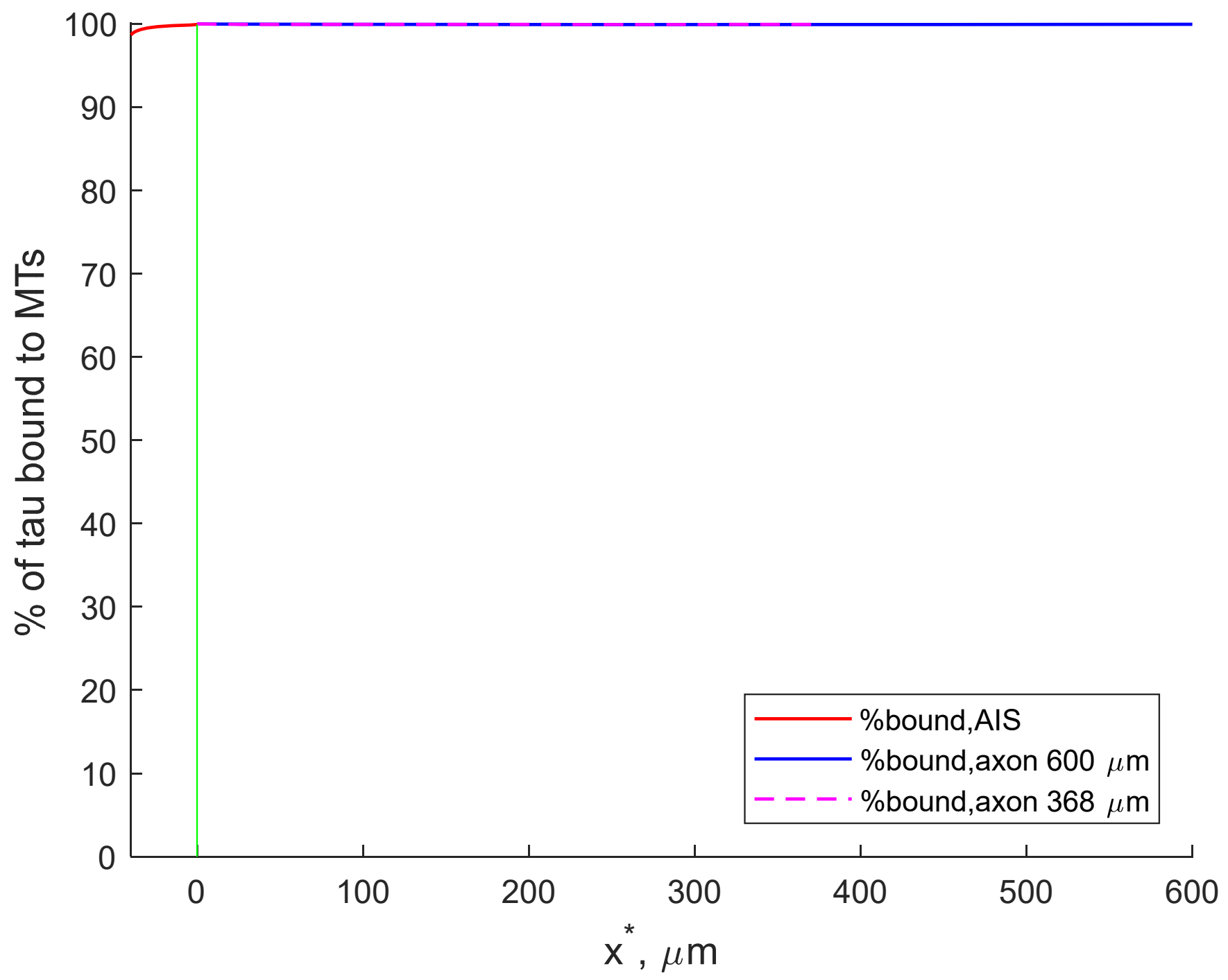

Fig. S3. (a) Concentration of tau diffusing along MTs, no association with motors. (b) Percentage of MT-bound tau, versus position in the axon. The vertical green line at $x$=0 shows the boundary between the AIS and the rest of the axon.

**S3.2. Tau concentrations in various kinetic states for two different values of the kinetic constant $\gamma_{on,a(1)}$**

(a)

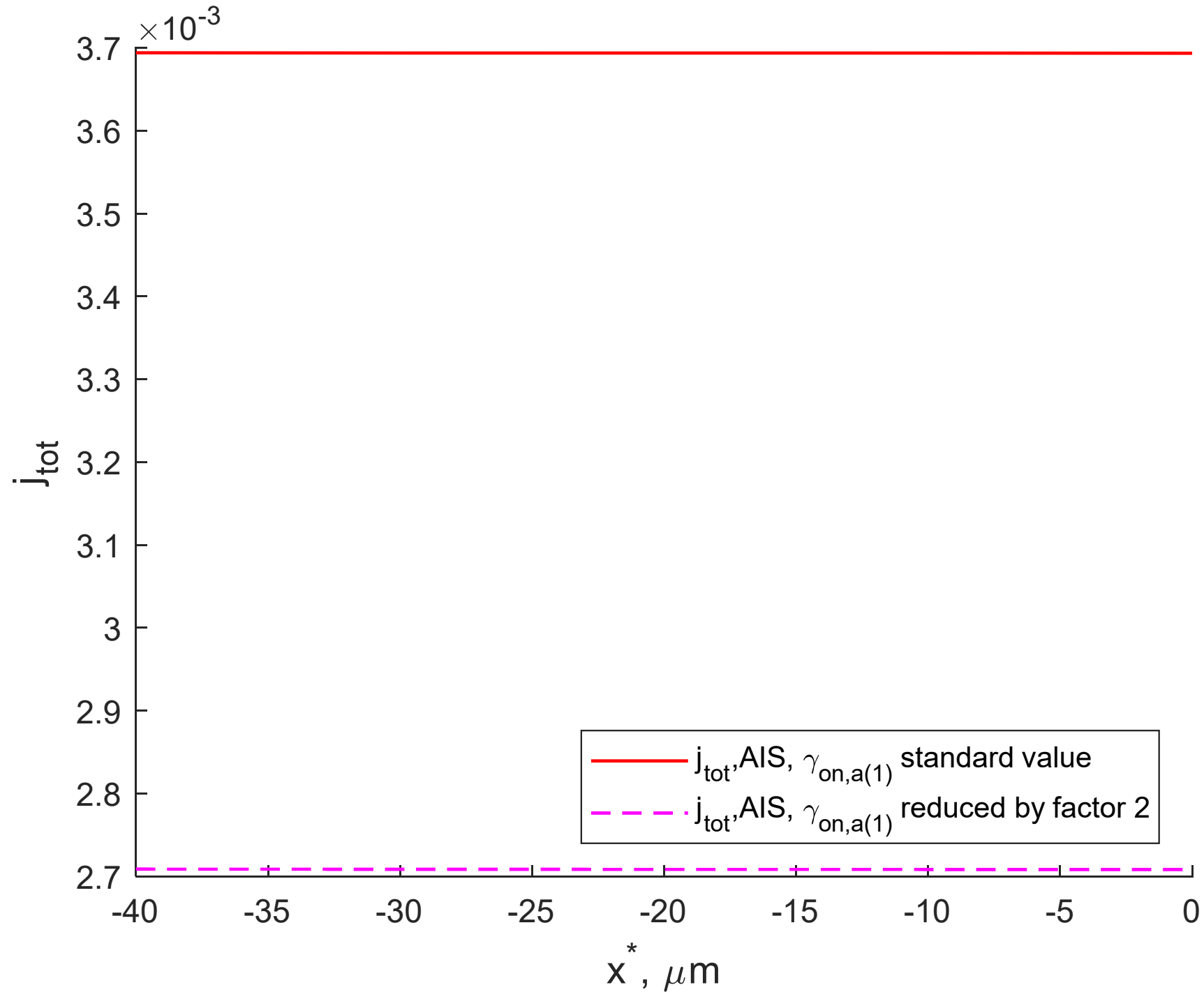

(b)

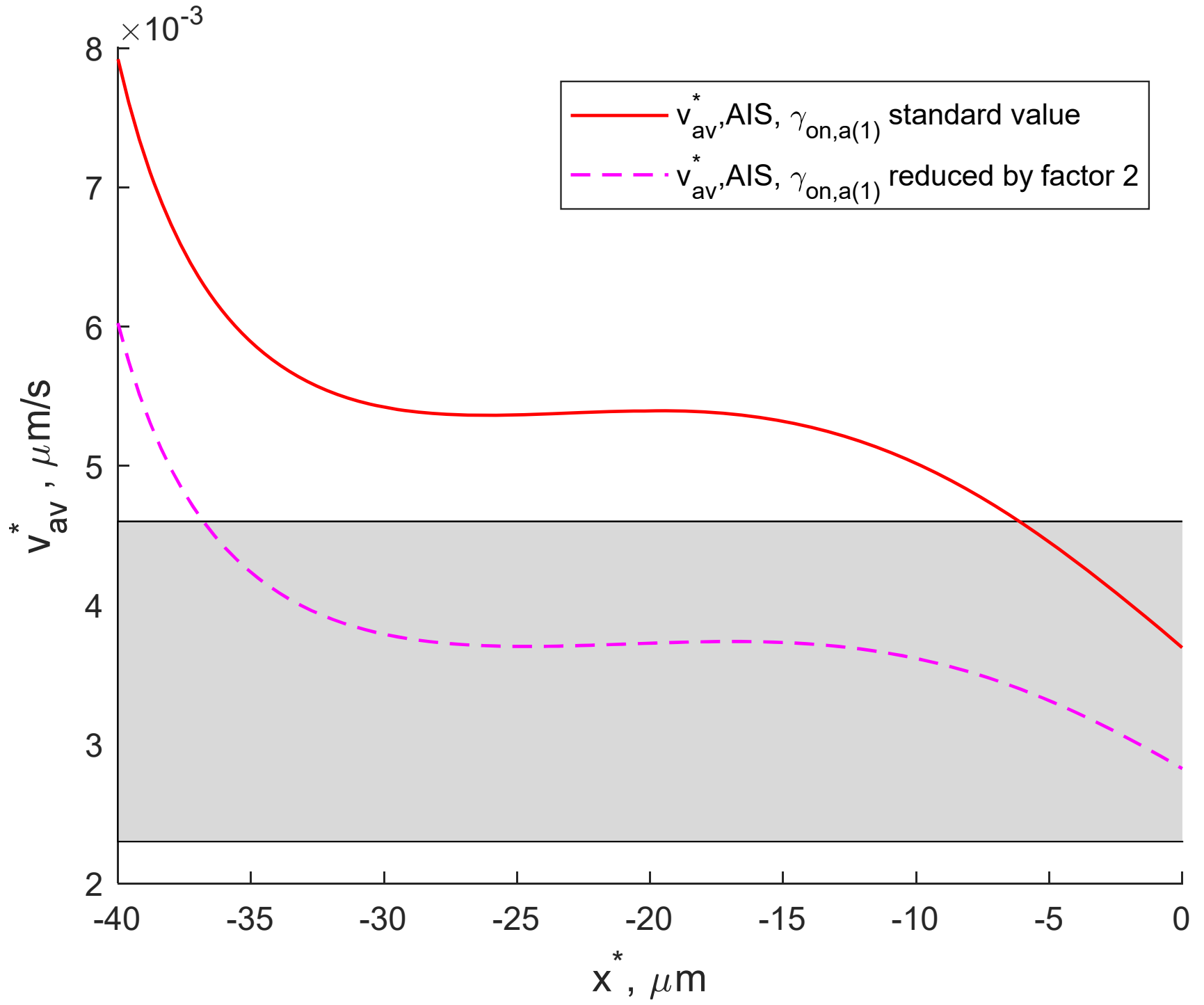

Fig. S4. Computations using the developed AIS model. (a) Total flux of tau. (b) Average velocity of tau, versus position in the axon. The range of the average tau velocity reported in [55] is shown by a horizontal band. Computations are carried out for two values of $\gamma_{on,a(1)}$ : a standard value reported in the fifth column of Table S2 and a value reduced by factor of two, $\gamma_{on,a(1)} / 2$ . Axon length was 600 μm.

(a)

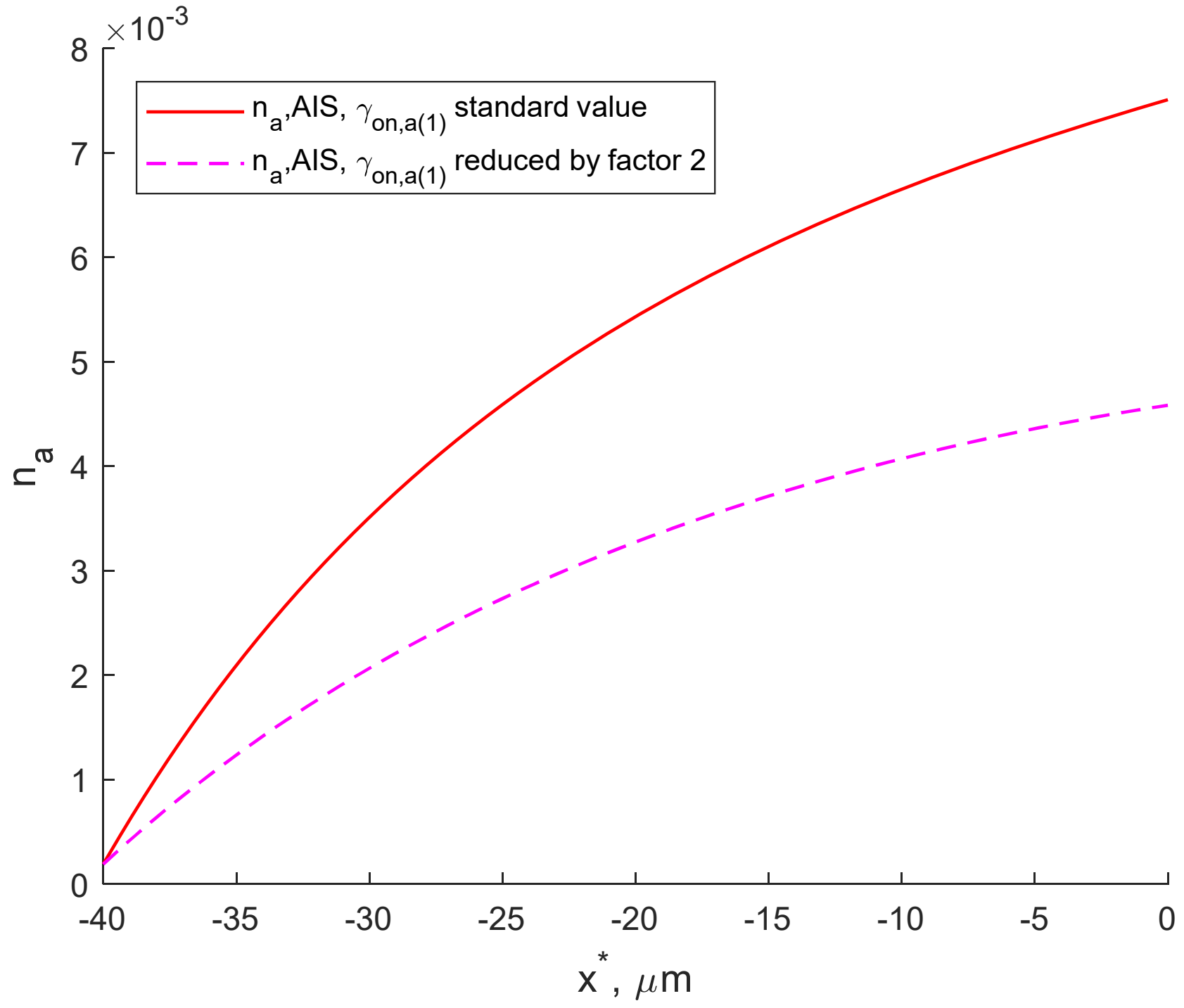

(b)

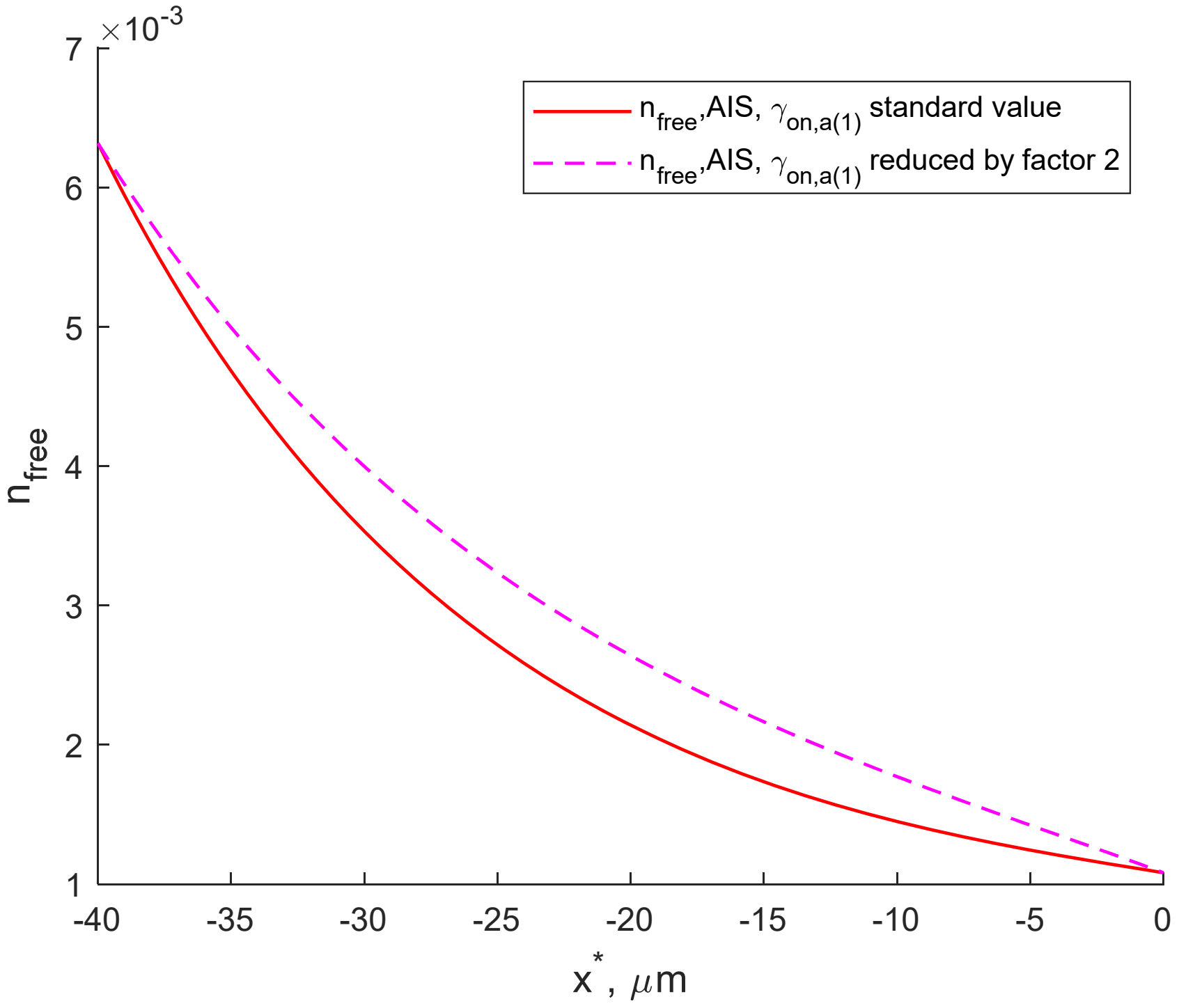

Fig. S5. Computations using the developed AIS model. (a) Concentration of on-track tau transported by molecular motors in the anterograde direction. (b) Concentration of free (diffusing) tau. Computations are carried out for two values of $\gamma_{on,a(1)}$ : a standard value reported in the fifth column of Table S2 and a value reduced by factor of two, $\gamma_{on,a(1)} / 2$ . Axon length was 600 μm.